\documentclass[acmsmall]{acmart}

\AtBeginDocument{%
  \providecommand\BibTeX{{%
    \normalfont B\kern-0.5em{\scshape i\kern-0.25em b}\kern-0.8em\TeX}}}

\usepackage{placeins}

\begin{document}

\received{May 13, 2025}
\received[revised]{January 13, 2026}
\received[accepted]{March 17, 2026}

\title{Investigating LLM-Powered Dissenting Minority Support in Power-Imbalanced Group Decision-Making: Counterargument and Mediation as Intervention Strategies}

\author{Soohwan Lee}
\orcid{0000-0001-8652-3408}
\affiliation{\institution{Department of Design, UNIST}
\city{Ulsan}
\country{Republic of Korea}}
\email{soohwanlee@unist.ac.kr}

\author{Seoyeong Hwang}
\orcid{0009-0004-1045-1419}
\affiliation{\institution{Department of Design, UNIST}
\city{Ulsan}
\country{Republic of Korea}}
\email{hseoyeong@unist.ac.kr}

\author{Mingyu Kim}
\orcid{0009-0006-8580-3532}
\affiliation{\institution{Department of Design, UNIST}
\city{Ulsan}
\country{Republic of Korea}}
\email{gyu7991@unist.ac.kr}

\author{Dajung Kim}
\orcid{0000-0002-9144-7435}
\affiliation{\institution{Department of Design, UNIST}
\city{Ulsan}
\country{Republic of Korea}}
\email{dajungkim@unist.ac.kr}

\author{Kyungho Lee}
\orcid{0000-0002-1292-3422}
\affiliation{\institution{Department of Design, UNIST}
\city{Ulsan}
\country{Republic of Korea}}
\email{kyungho@unist.ac.kr}

\begin{abstract}
Minority viewpoints are often suppressed in power-imbalanced group decision-making due to social pressure to comply with the majority. To address this problem, we developed an LLM-powered dissenting minority support system that aimed to foster attention to minority views through either AI-generated counterarguments or AI-mediated messages. We conducted a mixed-method experiment with 96 participants in 24 groups, comparing minority members' experiences across baseline, AI-counterargument, and AI-mediated message conditions. Our findings revealed a nuanced trade-off: AI-generated counterarguments fostered a more flexible atmosphere and enhanced satisfaction, while AI-mediated messaging increased minority participation but unexpectedly reduced their psychological safety. This research contributes empirical evidence on how different AI implementations affect group dynamics, identifies a critical support paradox between participation and psychological safety, provides design implications for future systems, and highlights ethical challenges in implementing AI-mediated communication in hierarchical settings. These insights advance understanding of designing more equitable AI support for power-imbalanced group decision-making.
\end{abstract}

\ccsdesc[500]{Human-centered computing~Computer supported cooperative work}
\ccsdesc[300]{Human-centered computing~Collaborative interaction}
\ccsdesc[300]{Human-centered computing~Natural language interfaces}
\ccsdesc[300]{Human-centered computing~HCI theory, concepts and models}

\keywords{group decision-making, conversational agents, critical thinking, social influence, LLM}

\begin{teaserfigure}
  \centering
  \includegraphics[width=1.0\textwidth]{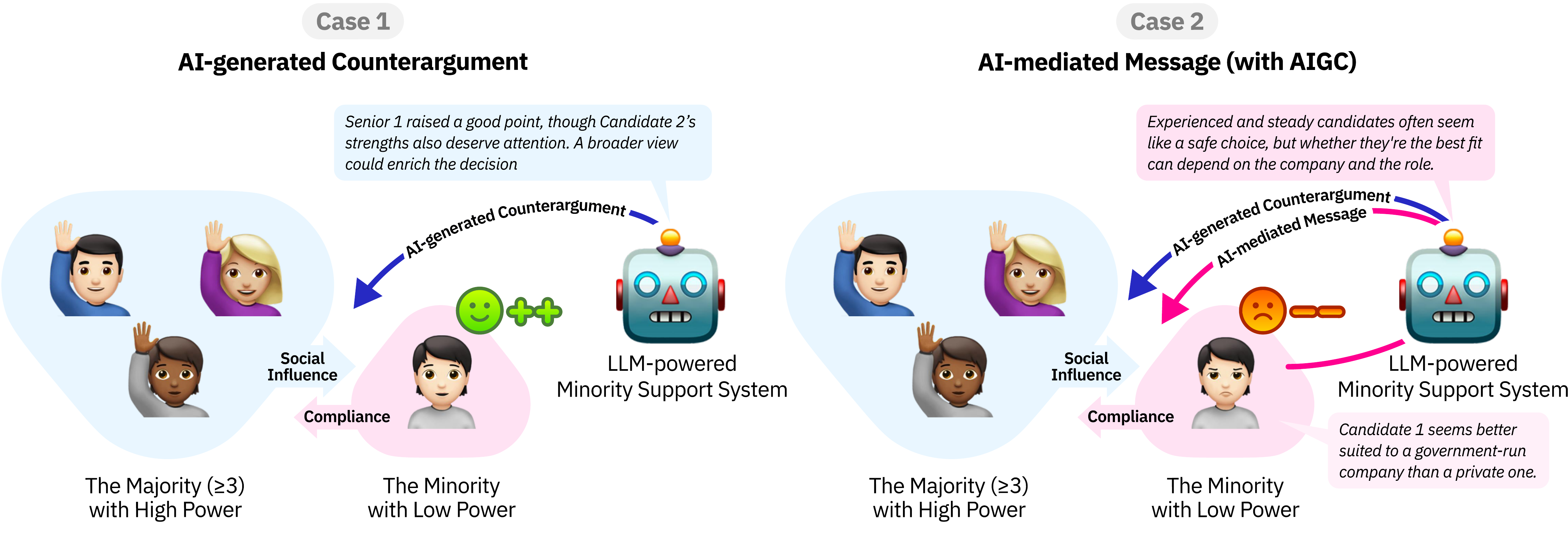}
  \caption{An LLM-powered dissenting minority support system mediates between majority and minority members through two designs. In AI-generated Counterargument (AIGC) condition, the system autonomously produces counterpoints to majority consensus, broadening discussion space and modeling dissent as conversational norm. In the AI-mediated Message (AIMM) condition, minority members can privately submit dissenting views that the system paraphrases and posts publicly, blended with AIGC outputs to mask authorship. While AIGC improved group atmosphere and satisfaction, AIMM increased minority participation but unexpectedly reduced their psychological safety and satisfaction.}
  \Description{The diagram shows an LLM-powered minority support system mediating between a majority group (≥3 members with high power, in blue) and a minority member (low power, in pink). The AI generates counterarguments (blue arrow) to challenge the majority and delivers supportive messages (pink arrow) to the minority to reduce isolation fear. The majority pressures the minority, leading to potential compliance (black arrow). This system aims to balance group dynamics and promote inclusive discussions.}
  \label{fig:teaser}
\end{teaserfigure}

\setcopyright{cc}
\setcctype{by}
\acmJournal{PACMHCI}
\acmYear{2026} \acmVolume{10} \acmNumber{6} \acmArticle{CSCW106}
\acmMonth{10} \acmDOI{10.1145/3816954}

\maketitle

\section{Introduction}
Power imbalances in group decision-making frequently suppress minority perspectives, restrict the diversity of ideas, and weaken overall outcomes \cite{janisGroupthinkPsychologicalStudies1982}. Compliance pressures often lead individuals to publicly align with the majority despite private disagreement, undermining psychological safety and discouraging meaningful participation from less-empowered members \cite{kelmanComplianceIdentificationInternalization1958, moscoviciStudiesSocialInfluence1976}. While collaborative processes help groups solve complex challenges, support individual learning, and lead to more accurate and creative outcomes across domains such as healthcare, education, and research \cite{maciejovskyTeamsMakeYou2013, voglerTeamBasedTestingImproves2016, glickInflictedTraumaticBrain2007, uzziAtypicalCombinationsScientific2013}, these benefits diminish when hierarchical structures or conformity suppress dissent. In an effort to address barriers to more balanced, active participation, recent studies have introduced AI-powered interventions that facilitate idea generation, consensus-building, and alternative viewpoint introduction to encourage engagement and mitigate groupthink \cite{kimBotBunchFacilitating2020, shinChatbotsFacilitatingConsensusBuilding2022, chiangEnhancingAIAssistedGroup2024}. These interventions typically operate through three underlying mechanisms: regulating atmosphere using affective cues to foster supportive group climate, balancing participation by prompting quieter members and curbing dominance, and diversifying perspectives by ensuring groups consider broader alternatives.

However, many AI-powered interventions assume equal standing and fair opportunities for participation. While some work has examined AI support under imbalanced group settings \cite{hwangSoundSupportGendered2024, liImprovingNonNativeSpeakers2022}, less explored is how AI can represent and route dissent when dissenters risk retaliation or identification, not just a lack of airtime. The existing approaches may improve aggregate engagement, but how they effectively address the challenge of supporting minority voices under conditions of unequal power remains underexplored. In many real-world settings, compliance emerges not only from numerical disadvantage but also from structural power asymmetry, such as when a junior member deliberates with multiple senior colleagues.\footnote{Throughout this paper, "minority" refers to participants' numerical and opinion-based position within group dynamics (consistent with Moscovici's minority influence theory \cite{moscoviciStudiesSocialInfluence1976}), specifically those who hold dissenting views and are outnumbered, not to demographic characteristics such as race, gender, or ethnicity.} Research on supporting and amplifying minority voices has also explored the potential of systematic interventions, such as anonymous feedback platforms and automated counterargument generators, to address these issues \cite{hwangSoundSupportGendered2024, liImprovingNonNativeSpeakers2022, duanIncreasingNativeSpeakers2019}. Yet, there is a significant gap, especially in how to represent authentic minority voices or adequately safeguard psychological safety and anonymity, which are crucial for minorities' experiences in group decision-making.

To address this gap, this research empirically investigates how LLM-powered minority support interventions can foster psychologically safe and equitable environments for minority members expressing dissent in hierarchical decision-making settings.

We designed two intervention conditions around how dissent enters the shared discussion. In the AI-generated Counterargument (AIGC) condition, inspired by prior work on devil's advocate systems \cite{chiangEnhancingAIAssistedGroup2024}, the AI periodically produces system-authored counterpoints to stimulate deliberation and broaden the option space. In the AI-mediated Message (AIMM) condition, we extend this approach by adding a private input channel for the minority member, then paraphrasing and injecting that input into the public thread while keeping AIGC active to mask authorship. This coupling is intentional: by sustaining a stream of AI-authored counterarguments, AIMM makes it difficult for others to infer whether any given system message is "purely AI-generated" or "revoiced from a member," thereby aiming to preserve anonymity through indistinguishability rather than through explicit anonymization alone. We expected that AIMM, by concealing even the presence of dissent within the group, would offer stronger psychological safety than AIGC alone, enabling minority members to participate more actively and feel greater satisfaction under compliance pressure. We examine the influence of these conditions on group decision making under compliance pressure with four factors. To what extent do these conditions affect \textbf{Perceived psychological safety (RQ1)}, \textbf{Engagement (RQ2)}, \textbf{Satisfaction with decision-making processes and outcomes (RQ3)}, and \textbf{Cognitive workload of participants in hierarchical discussions (RQ4)}?

To investigate these questions, we conducted a mixed-method study with 96 participants organized into 24 groups of four members each. Each group consisted of three members assigned as high-power majority (seniors) and one member assigned as low-power minority (junior), with roles randomly assigned. We employed a mixed experimental design with participant type (majority vs. minority) as a between-subjects variable and type of LLM-powered minority support (AIGC, AIMM) as a partially within-subjects variable. 
Our findings indicate that AI-generated counterarguments (AIGC) improved group atmosphere and participant satisfaction. However, enabling minority members to submit opinions anonymously via AI (AIMM) increased discussion while simultaneously reducing psychological safety and satisfaction for minority participants.
Notably, AIMM revealed a paradoxical pattern: provenance-cloaked revoicing can increase participation while simultaneously eroding psychological safety, suggesting that anonymity-by-indistinguishability and message-level proxying are not inherently empowering in hierarchical discussions. These results underscore the complex trade-offs between increasing minority participation and maintaining psychological safety when using LLM-powered support for minority in group decision-making.

This research makes four key contributions to CSCW. First, we empirically distinguish how two LLM-powered intervention strategies—AI-generated counterarguments (AIGC) versus revoicing-based assistant-mediated mediation (AIMM)—shape the experiences of minority members under compliance pressure in power-imbalanced group decision-making. Second, we identify a critical \emph{support paradox}: while AIMM can increase minority participation, it can simultaneously decrease psychological safety and satisfaction, challenging the assumption that mediated anonymity straightforwardly protects minority members. Third, we derive actionable design implications for AI support in hierarchical discussions, providing evidence-informed guidance on when systems can introduce dissent as independent counterarguments and when revoicing participant input may be less appropriate, highlighting the role of agency, transparency, and perceived authorship. Finally, we discuss ethical considerations and deployment risks of revoicing-based interventions in organizational settings, where AI-mediated communication may unintentionally intensify marginalization by discounting or obscuring human dissent.

\section{Related Work}
\subsection{The Impact of Social Influence and Power on Group Decision-making}
Group decision-making leverages collective intelligence to produce superior outcomes across various domains \cite{maciejovskyTeamsMakeYou2013,glickInflictedTraumaticBrain2007,uzziAtypicalCombinationsScientific2013}, but these processes are significantly shaped by social influence and power dynamics \cite{moscoviciStudiesSocialInfluence1976,kelmanComplianceIdentificationInternalization1958}. Social influence theory suggests that individuals tend to adjust their behavior to meet social demands, with the majority's opinions exerting particularly strong pressure on those with less power in the group. Moscovici's conversion theory specifically explains that majority influences trigger a comparison process resulting in compliance - a form of conformity where individuals outwardly agree while maintaining private disagreement \cite{moscoviciStudiesSocialInfluence1976}. This compliance is typically direct, immediate, and temporary, serving as a coping mechanism in power-imbalanced situations rather than reflecting genuine belief change.

Power dynamics become especially problematic in hierarchical settings where power imbalances are formalized through reward and legitimate power structures \cite{frenchjr.BasesSocialPower1959}. Kelman's framework provides particular insight here, identifying compliance as an initial response to power, in which individuals conform primarily to avoid repercussions or gain rewards, rather than from genuine conviction \cite{kelmanComplianceIdentificationInternalization1958}. This dynamic is especially evident among minority members, who are often treated as outgroup members and experience isolation. The effect is particularly pronounced when the size disparity between majority and minority groups is substantial. The resulting self-censorship triggers a cascade of negative effects: as minority voices are silenced, groups lose access to diverse perspectives that could enhance decision quality, ultimately leading to groupthink, where the desire for consensus overrides critical evaluation of alternatives \cite{janisGroupthinkPsychologicalStudies1982,janisVictimsGroupthinkPsychological1972,kamedaPsychologicalEntrapmentGroup1993}.

Traditional social psychology offers several approaches to address compliance and prevent its progression to groupthink in group settings.  The devil's advocate technique addresses groupthink by assigning someone to challenge prevailing viewpoints, stimulating opinion diversity  \cite{macdougallDevilsAdvocateStrategy1997,masonDialecticalApproachStrategic1969,nemethDevilsAdvocateAuthentic2001,schweigerGroupApproachesImproving1986,schwenkEffectsDevilsAdvocacy1994}. However, this method doesn't directly solve the underlying compliance issues and faces significant limitations: designated advocates often lack authentic dissenting perspectives, risk social ostracism when challenging powerful members, and cannot accurately represent unexpressed minority viewpoints \cite{nemethDevilsAdvocateAuthentic2001,schulz-hardtProductiveConflictGroup2002,jamiesonSympathyDevilPhysiological2014}. Other compliance-reduction strategies include leadership interventions to create psychologically safe environments \cite{edmondsonPsychologicalSafetyLearning1999} and anonymous feedback channels \cite{jessupEffectsAnonymityGDSS1990a}. However, these methods face practical challenges: leadership interventions rely on the leader's skills, and anonymous feedback often fails in small groups where unique opinions reveal identities. This research addresses these limitations by exploring how conversational AI might provide both psychological safety and true anonymity while preserving the benefits of diverse perspectives in decision-making processes.

\subsection{AI-powered Approaches to Improving Group Decision-Making}

HCI research on AI-assisted decision-making initially focused on supporting individuals, where AI provides recommendations, rationales, or safeguards while people retain final authority \cite{laiScienceHumanAIDecision2023a, bucincaTrustThinkCognitive2021, yataniAIExtrahericsFostering2024,maWhoShouldTrust2023,swaroopPersonalisingAIAssistance2025,munyakaDecisionMakingStrategies2023}. As collaborative work increasingly unfolds through online discussions, recent systems shift attention to \emph{group} decision-making, where outcomes depend on interaction patterns, participation, and social influence rather than on isolated judgments \cite{maRecommenderExploratoryStudy2024, seboRobotsGroupsTeams2020, zhangBreakingBarriersBuilding2025, kimBotBunchFacilitating2020,chiangAreTwoHeads2023}. In this space, prior work implicitly assigns AI different roles in shaping the discussion. We organize this landscape around four recurring roles—moderation, facilitation, mediation, and debating—distinguished by what aspect of group interaction the AI primarily intervenes in.

Moderation and facilitation focus on the process of discussion but operate at different levels. Moderation enforces conversational boundaries to ensure safety and trustworthiness, for example, by filtering harmful content, controlling off-topic drift, or enforcing participation rules \cite{agarwalConversationalAgentsFacilitate2024, leeCounteringMisinformationPrivate2025,doanDesignSpaceOnline2025}. Facilitation instead supports the flow and continuity of interaction by scaffolding turn-taking, prompting quieter members, and sustaining goal-oriented discussion \cite{houdeControllingAIAgent2025, heAIFutureCollaborative2024, doHowShouldAgent2022, doInformExplainControl2023, kimBotBunchFacilitating2020}. While both roles optimize group dynamics, they typically assume relatively equal standing among participants and treat disagreement as a coordination issue rather than a social risk. Mediation occupies a different position. Rather than a governing process, mediation lightly bridges perspectives by reframing or contextualizing social signals to reduce misunderstanding and friction \cite{phamEmbodiedMediationGroup2024, doanDesignSpaceOnline2025, goversAIDrivenMediationStrategies2024,claggettRelationalAIFacilitating2025}, or facilitating group consensus \cite{shinChatbotsFacilitatingConsensusBuilding2022, tesslerAICanHelp2024a}, also often including AI-mediated communication \cite{fuTextSelfUsers2024, hancockAIMediatedCommunicationDefinition2020}. These approaches do not claim the neutrality of classical dispute mediation, but they raise questions about agency, attribution, and ownership that we address in the following subsection. Debating-oriented systems intervene at the level of reasoning by contributing system-authored opinions \cite{fulayEmptyChairUsing2025,leongDittosPersonalizedEmbodied2024,zhengCompetentRigidIdentifying2023} or Socratic-style questions that challenge emerging consensus and prompt critical reflection \cite{chiangEnhancingAIAssistedGroup2024,parkThinkingAssistantsLLMBased2024}.

Across moderation, facilitation, mediation, and debating, however, many systems relatively under-specify how power imbalance and conformity pressure shape participation \cite{hwangSoundSupportGendered2024, liImprovingNonNativeSpeakers2022, duanIncreasingNativeSpeakers2019,gaoEffectsPublicVs2014,gaoImprovingMultilingualCollaboration2015}, even though social psychology shows that minorities often comply publicly despite private disagreement \cite{forsythGroupDynamics2018, kelmanComplianceIdentificationInternalization1958, moscoviciStudiesSocialInfluence1976, janisGroupthinkPsychologicalStudies1982}. Our work fills this gap by examining LLM-powered support under explicit hierarchical pressure. We compare two intervention logics that cut across this taxonomy: AIGC instantiates a debating role through system-authored counterarguments that normalize dissent, while AIMM combines facilitation and light mediation by allowing minorities to route dissent through the system without presenting the AI as an independent mediator. This framing clarifies how different AI roles interact with power and psychological safety in group decision-making.

\subsection{AI-Mediated Communication for Group Interaction}
AI-mediated communication (AIMC) encompasses scenarios where "a computational agent operates on behalf of a communicator by modifying, augmenting, or generating messages to accomplish communication or interpersonal goals" \cite{hancockAIMediatedCommunicationDefinition2020}. Prior AIMC research shows AI can generate content on request \cite{hancockAIMediatedCommunicationDefinition2020}, suggest communication strategies \cite{doHowShouldAgent2022}, rephrase messages for tone \cite{fuTextSelfUsers2024}, or facilitate perspective-sharing in groups \cite{wangUnderstandingDesignSpace2022, shinChatbotsFacilitatingConsensusBuilding2022}. However, these systems primarily support individual communication, such as drafting emails or refining text, without addressing how power imbalances silence minority voices in group settings. AIMC can reconfigure group dynamics through communication patterns, tone, and interpersonal interactions \cite{mieczkowskiAIMediatedCommunicationLanguage2021, poddarAIWritingAssistants2023, hohensteinArtificialIntelligenceCommunication2023, robertsonCantReplyThat2021}, but its potential to reshape power structures remains relatively underexplored.

While Shin et al. explored consensus-building through asynchronous AIMC \cite{shinChatbotsFacilitatingConsensusBuilding2022}, they did not address minority protection under power imbalances. Our work extends AIMC by comparing two interventions for power-imbalanced group decision-making: AIGC (AI-Generated Counterarguments), where AI autonomously posts counterpoints, and AIMM (AI-Mediated Messages), where AI paraphrases minority input and posts it as its own. AIGC acts as a Devil's Advocate that normalizes dissent without revoicing human input. \textbf{AIMM performs limited mediation by revoicing minority input to mask authorship, without facilitating dialogue or resolving disputes like traditional mediation \cite{hancockAIMediatedCommunicationDefinition2020,fuTextSelfUsers2024,stauferSilencingRiskNot2024}.} This narrow focus on anonymity-through-revoicing addresses a critical limitation: traditional anonymous channels reveal that dissent exists within the group \cite{jessupEffectsAnonymityGDSS1990a, stauferSilencingRiskNot2024}. \textbf{By combining revoicing with AIGC's autonomous counterarguments, AIMM makes messages indistinguishable in provenance, preserving complete anonymity while providing minority perspectives.} However, this design trades authenticity and attribution for anonymity, raising questions about authorship, agency, and legitimacy. By comparing AIGC and AIMM, we investigate how different AI roles affect minority experiences under compliance pressure, extending established AIMC frameworks \cite{hancockAIMediatedCommunicationDefinition2020, wangUnderstandingDesignSpace2022, fuTextSelfUsers2024} to power-imbalanced collaborative decision-making.

\section{System Design \& Implementation: LLM-powered Minority Support System}

\subsection{System Concept and Design Rationale}

\subsubsection{System Concept and Working Definition}

This study presents two AI tools that help minority voices in group decisions: (1) an AI-generated Counterargument (AIGC) system that creates opposing views to expand discussion, and (2) an AI-mediated Message (AIMM) system that shares minority members' ideas anonymously while also creating counterarguments. These ideas come from traditional devil's advocate methods, which aim to encourage reflection by introducing disagreement. However, these methods often become fake or isolate the person assigned to disagree \cite{nemethDevilsAdvocateAuthentic2001,schulz-hardtProductiveConflictGroup2002,jamiesonSympathyDevilPhysiological2014}. Also, they do not give minorities a safe way to communicate, especially in groups with power imbalances, highlighting the need for alternative designs that both expand discussion and provide secure channels for dissent.

We frame both AIGC and AIMM under the broader concept of an LLM-powered Minority Support System. This system is a real-time conversation agent that joins group discussions with two goals: (a) expanding group thinking through counterarguments and (b) protecting minority voices through anonymous sharing. In the AIMM condition, keeping the counterargument function is important. Without it, the system would look like a simple anonymous suggestion box, making it clear that someone in the group disagrees, which can be dangerous in closed or conformist settings. To preserve anonymity, the agent intentionally withholds the provenance of its utterances, so the majority cannot tell whether any given message is an autonomous counterargument or a revoiced minority input. By combining counterargument creation with message sharing, AIMM makes minority dissenting input look the same as the agent's own ideas, creating better anonymity. When no minority input exists, the agent summarizes current opinions and asks questions that invite different perspectives. This definition applies to our specific design, where the system acts as an interactive participant in group talk that maintains critical thinking while protecting vulnerable voices.

\subsubsection{Design Scope and Rationale}
The LLM's counterargument generation and paraphrasing capabilities are critical technical components in this system. Our research focuses primarily on the social and psychological impacts of introducing such a system into group decision-making processes. Building on this scope, we structured the AI's counterargument approach based on previous research in AI-assisted decision-making. We implemented five key design considerations to maximize effectiveness. While the majority of these design features apply to both AIGC and AIMM, the anonymous revoicing mechanism is specific to AIMM. First, we generated feedback at regular eight-turn intervals to maintain engagement without overwhelming participants, allowing natural conversation flow. The eight-turn interval was chosen to allow each participant in our four-person groups to make at least two comments—one expressing their own opinion and one responding to another's—before intervention. Second, we designed the system to ask questions rather than provide direct logical counterarguments, as research indicates this approach more effectively prompts participants to think critically \cite{danryDontJustTell2023, yataniAIExtrahericsFostering2024}. Third, we employed persuasive rather than confrontational rhetoric, as persuasive language better promotes critical thinking in AI-human interactions \cite{tanprasertDebateChatbotsFacilitate2024}. Fourth, we deliberately avoided repeating previous statements to prevent redundancy when group opinions remained static during short decision-making sessions with no direct AI interaction \cite{milanaChatbotsAdvisersEffects2023, xuetaoImpactAgentsAnswers2009}. Finally, in the AIMM condition, the system presents paraphrased human input as its own opinions, creating a fully anonymous channel for minority viewpoints while protecting contributor identity. These considerations work together to promote critical thinking and preserve anonymity throughout group discussions.

\begin{figure*}[]
  \centering
  \includegraphics[width=1.0\textwidth]{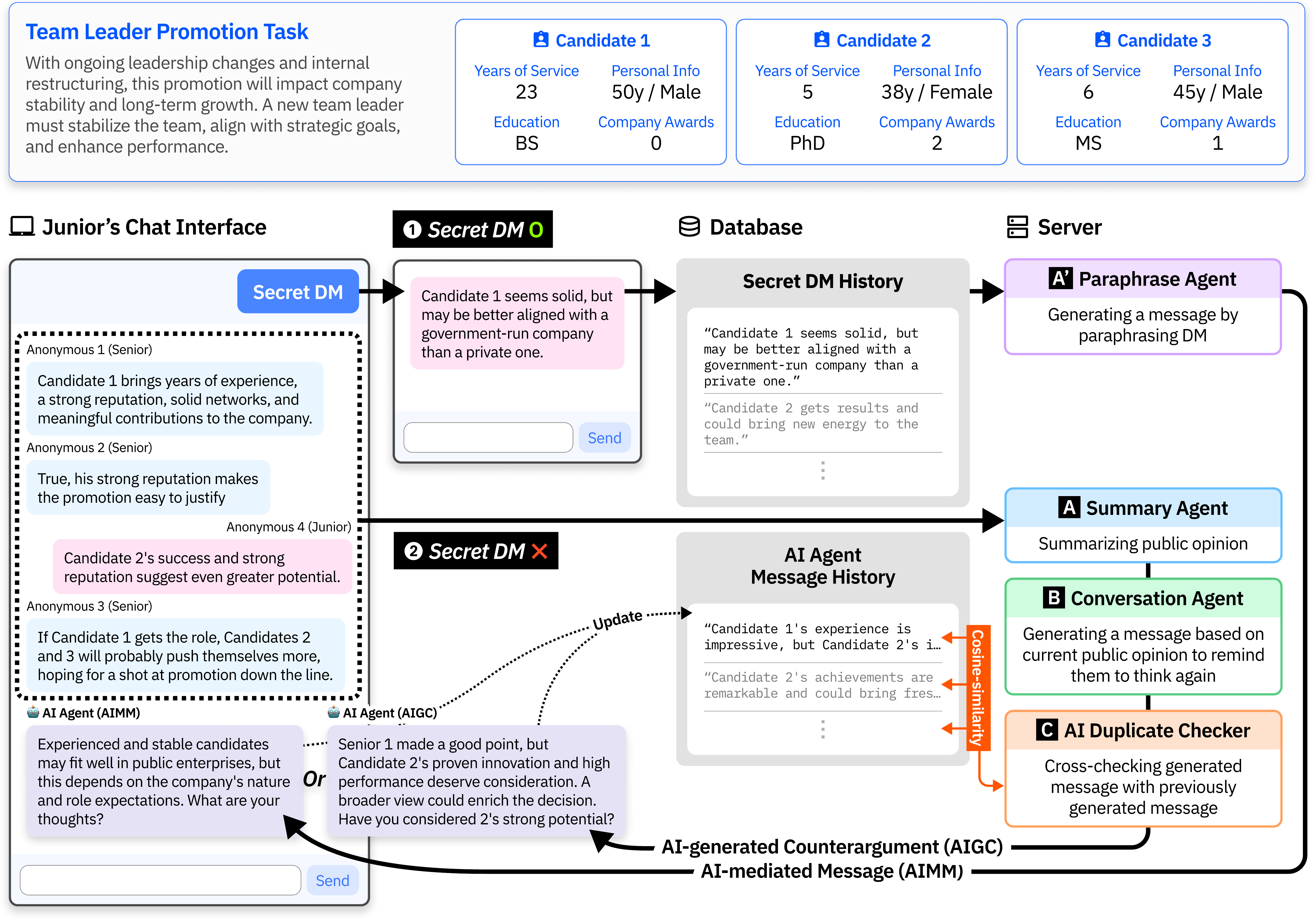}
  \caption{System Overview and Example Task Scenario. The figure illustrates a team leader promotion decision task, where participants discuss candidate qualifications in a chat interface. Minority members can privately share dissenting views via direct messages (DM) to the system, which reformulates and presents them as AI-mediated messages. If there is no DM with an opposing opinion, the system will send out a counterargument that it has generated on its own. The system architecture consists of a chat interface, database, and server, processing both public discussions and private DMs through four key agents: (A) Summary Agent for analyzing public opinion, (A') Paraphrase Agent for rephrasing minority views, (B) Conversation Agent for generating contextual counterarguments, and (C) AI Duplicate Checker for ensuring message novelty via cosine-similarity comparison.}
  \Description{This figure illustrates a system for group decision-making in a team leader promotion task, where participants discuss candidate qualifications in a chat interface. Minority members can privately share dissenting views via direct messages (DMs) to the AI, which reformulates and presents them as AI-generated insights to reduce social pressure. The system consists of a chat interface, database, and server with four AI agents: (A) Summary Agent for analyzing public opinion, (A’) Paraphrase Agent for anonymizing minority views, (B) Conversation Agent for generating counterarguments, and (C) AI Duplicate Checker for ensuring message novelty. If no dissenting DM is provided, the AI generates its own counterargument, fostering diverse perspectives and mitigating conformity bias in group discussions.}
  \label{fig:systemImplementation}
\end{figure*}

\subsection{System Architecture and Implementation}
We developed a custom online chat environment to enable integration of an LLM-powered Devil’s Advocate agent and to conduct controlled group discussions. The frontend uses TypeScript (React) and the backend uses Python (FastAPI). The LLM (OpenAI GPT-4o) interfaces with system modules, with Retrieval-Augmented Generation used only for referencing and paraphrasing direct messages sent to the LLM-powered Devil's Advocate.

Drawing on findings that LLMs often struggle to access mid-conversation information \cite{liuLostMiddleHow2023}, we employ a multi-agent architecture to clearly detect the majority opinion and encourage constructive discourse (\autoref{fig:systemImplementation} \& \autoref{Appendix-prompt}):
\textbf{(A) Summary Agent} – Consolidates emerging majority opinion to overcome LLM limitations in retaining mid-dialogue content \cite{liuLostMiddleHow2023}.
\textbf{(A') Paraphrase Agent} – Responds exclusively to direct messages from juniors, rearticulating their dissenting views as though originating from the AI itself. These messages are stored in a database with an \textit{"isUsed"} property; the agent retrieves entries where \textit{"isUsed"} is \textit{false}, sets it to \textit{true}, paraphrases the content, and outputs it as system-generated text.
\textbf{(B) Conversation Agent} – Encourages alternative perspectives by first empathizing with the other person’s point of view and then offering a gentle counterargument using a Socratic style.
\textbf{(C) AI Duplicate Checker} – Identifies repetitive content by calculating semantic similarity between sentence embeddings generated using the ‘paraphrase-multilingual-MiniLM-L12-v2’ model on an NVIDIA A6000.

\section{Methods}
The purpose of this study was to investigate the influence of the two AI interventions (AIGC and AIMM) on psychological safety and satisfaction of low-power minority in power-imbalanced group-decision making.
To simulate situations where a low-power minority member experiences the pressure to comply with majority opinions, we asked one of the participants from each group to play a role as a Junior member in the group,  representing their low-power minority positions, whereas the other participants were asked to play a Senior role, representing their high-power majority positions.
We compare two AI-supported interventions—AI-Generated Counterarguments (AIGC) and AI-Mediated Messages (AIMM)—against a baseline condition to assess their effectiveness. Measurement evaluates psychological safety, engagement levels, decision quality perceptions, cognitive workload, and perception of AI across experimental conditions.

\begin{figure*}[]
  \centering
  \includegraphics[width=1.0\textwidth]{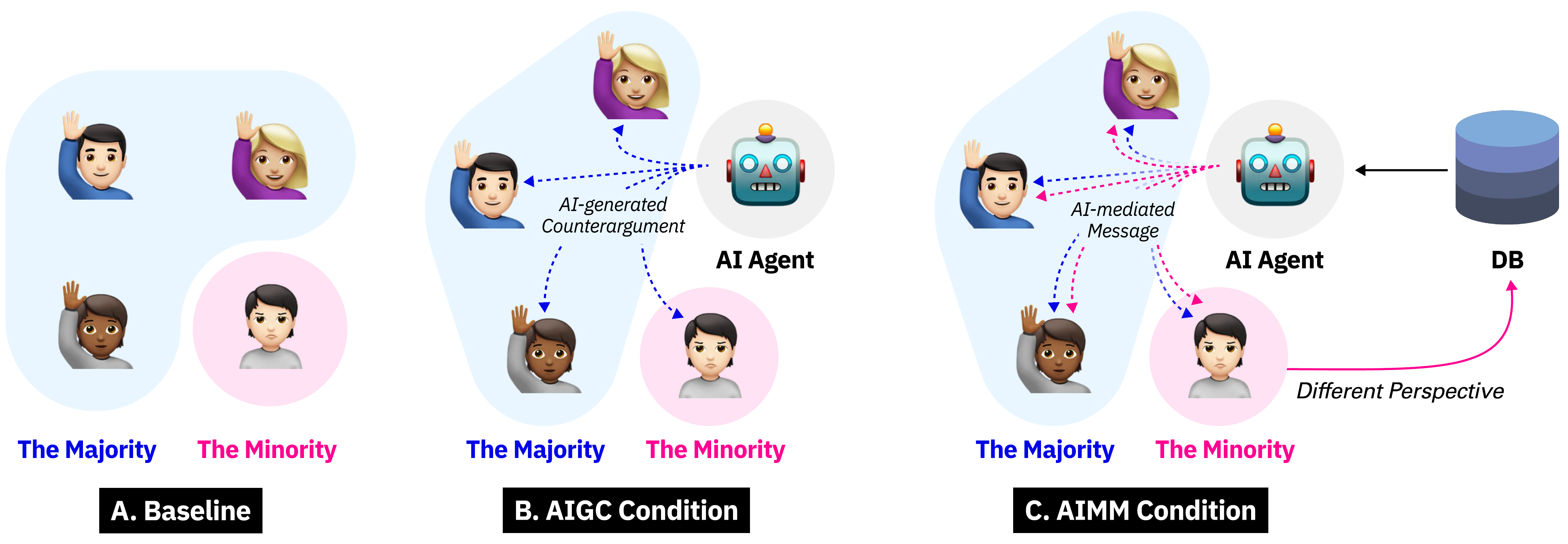}
  \caption{Experimental Conditions: Baseline shows the baseline group chat configuration with majority (blue) and minority (pink) participants. AIGC introduces an AI-powered minority support system that generates rebuttals during group discussions. AIMM extends this by enabling the minority member to privately send counterarguments to the AI system, incorporating them into its responses while maintaining anonymity.}
  \Description{The diagram presents three experimental conditions: Baseline: Baseline configuration where the majority (blue) and the minority (pink) interact directly in a group chat without AI intervention. AIGC: The LLM-powered Devil's Advocate introduces AI-generated counterarguments (blue arrows) to challenge majority views during discussions. AIMM: The Devil's Advocate integrates anonymous counterarguments privately sent by the minority (pink arrow to the AI), stored in a database (DB), and incorporated into AI-generated responses (blue arrows), ensuring the minority’s anonymity while influencing group discourse.}
  \label{fig:conditions}
\end{figure*}

\subsection{Experiment Conditions}

This study examines how different system conditions and participant types affect group communication under compliance pressure. Each participant experienced one of two system conditions: the AI-generated Counterargument condition or AI-mediated Message condition, alongside a common Baseline condition. Participants were randomly assigned to a role within each group: the Majority with High Power (Seniors) or the Minority with Low Power (A Junior).

\subsubsection{System Conditions}
To investigate how AI intervention shapes group dynamics, we designed three system conditions (\autoref{fig:conditions}):

\begin{itemize}
    \item \textbf{\textit{Baseline}}: A standard group chat setting without AI involvement. This condition served as the control for natural group discussion.
    \item \textbf{\textit{AI-Generated Counterargument (AIGC) Condition}}: This condition introduced an LLM-powered intervention that periodically and automatically generated counterarguments during the group discussion, inspired by the concept of Devil’s Advocate. The AI functioned independently, without access to private user input. The goal was to evaluate the pure effect of AI-led critical questioning on group discourse, separate from any anonymity or revoicing mechanisms, because the system design claims to be a devil's advocate. This condition reflects the core concept of a Devil’s Advocate as a neutral agent that challenges group consensus.
    \item \textbf{\textit{AI-Mediated Message (AIMM) Condition}}: This condition introduced AIMM, which combined AI-generated counterarguments with mediated messaging. Unlike AIGC, where AI autonomously generated dissenting points, AIMM allowed the minority member to submit input that AI would paraphrase and post as its own. \textbf{By blending revoiced minority input with AIGC counterarguments, AIMM makes messages indistinguishable in provenance—majority members cannot tell whether any AI message is autonomous or revoiced from the minority.} This addresses a critical limitation of traditional anonymous channels, which reveal that dissent exists within the group even when hiding individual identities. Through this mechanism, we aimed to protect minority members from hierarchical risks while preserving complete anonymity and encouraging active participation in group deliberation. Only the minority participant knew about this feature because our study focused on how minorities experience and use such support under compliance pressure.

\end{itemize}

To examine the unique effect of the AIMM condition, it was necessary to also test the AIGC condition alongside baseline; by comparing across all three, we could disentangle the impact of AI acting merely as a generalized Devil’s Advocate from the additional benefits of mediating minority voices through private, anonymous revoicing. To avoid revealing the experimental intent, each participant experienced only two of the three conditions: the baseline and one of the system conditions (AIGC or AIMM). This design minimized the risk that minority participants would recognize the specific purpose of the study or become aware of systematic differences between the two AI conditions, which could have altered their behavior.

\begin{figure*}[]
  \centering
  \includegraphics[width=1.0\textwidth]{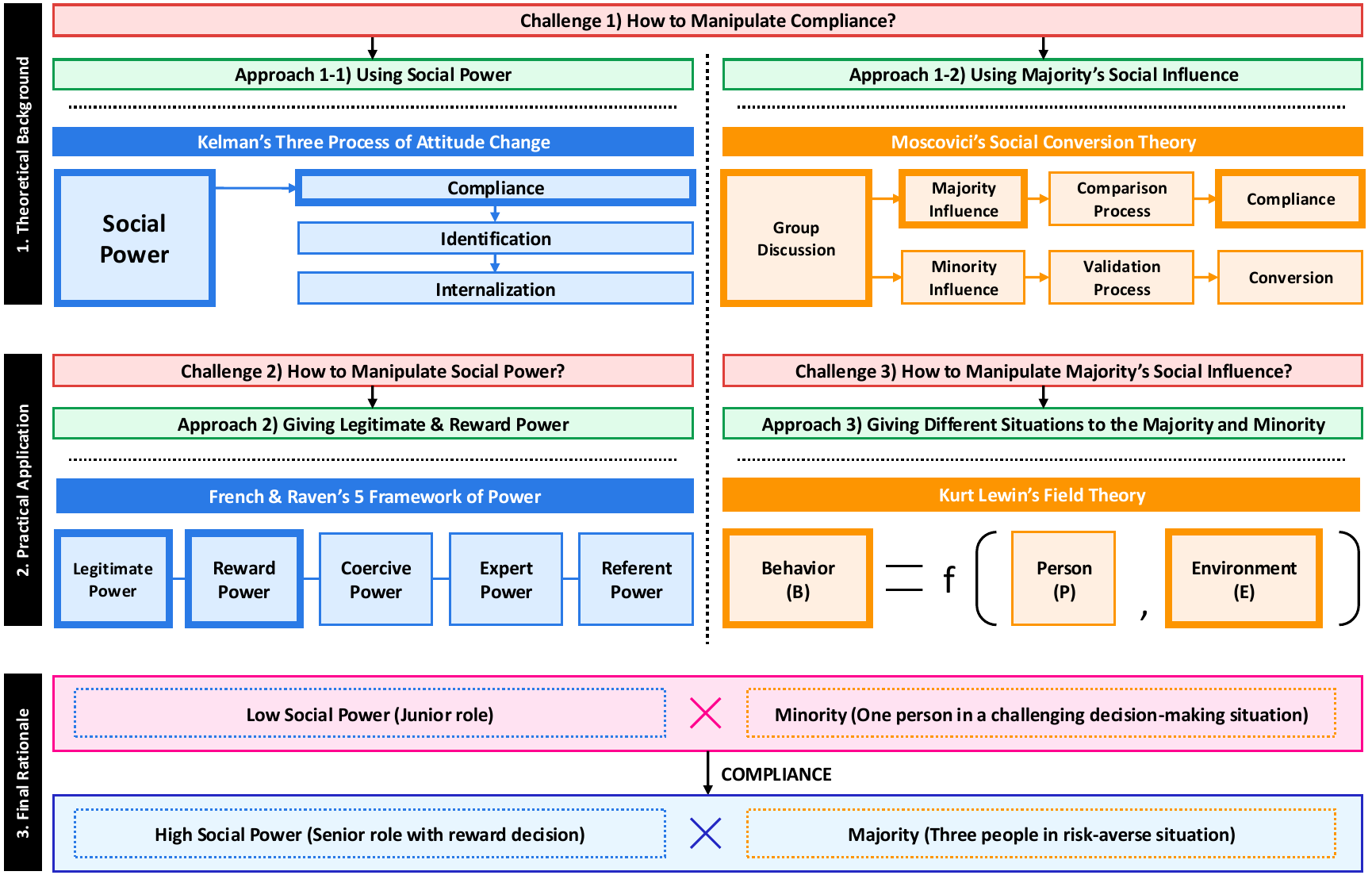}
  \caption{Theoretical Framework for Manipulating Compliance: This diagram illustrates our two-pronged approach to inducing experimental compliance: through social power (left) and the majority's social influence (right). The framework progresses from theoretical foundations (Kelman's and Moscovici's theories) to practical implementation (using legitimate/reward power and situational contexts), resulting in an experimental setup where a low-power minority (Junior) is positioned against high-power majorities (Seniors) to elicit compliance.}
  \Description{--}
  \label{fig:complianceFramework}
\end{figure*}

\subsubsection{Participant Types}
We created controlled compliance situations by manipulating two key mechanisms that induce compliance: social power and majority social influence (\autoref{fig:complianceFramework}).

Following Kelman's Theory of Attitude Change \cite{kelmanComplianceIdentificationInternalization1958}, which conceptualizes compliance as behavioral change driven by social power without internal acceptance (\autoref{fig:complianceFramework} - Approach 1-1), we operationalized social power using French and Raven's five bases of power \cite{frenchjr.BasesSocialPower1959}. These bases describe different sources through which influence operates in social relationships, including legitimate authority, control over rewards, capacity for coercion, perceived expertise, and interpersonal identification (\autoref{fig:complianceFramework} - Approach 2). Among these, we focused on legitimate and reward power by establishing a senior and junior hierarchy, as validated in prior work \cite{houShouldFollowHuman2023,houPowerHumanRobotInteraction2024}. Seniors received the authority to allocate additional compensation, while Juniors were informed that their compensation depended on Senior evaluations, creating a realistic context in which expressing dissent involved personal risk.

Majority social influence was implemented following Moscovici's Social Conversion Theory \cite{moscoviciStudiesSocialInfluence1976}. This theory shows that numerical majorities exert conformity pressure on minority members even in the absence of direct coercion (\autoref{fig:complianceFramework} - Approach 1-2). To induce stable majority and minority positions in an online setting with arbitrary decision-making tasks, we adopted a role-playing approach in which participants received different contextual information that guided them toward divergent initial preferences \cite{lewinFieldTheoryExperiment1939} (\autoref{fig:complianceFramework} - Approach 3). We employed a 3:1 majority-to-minority ratio, based on prior findings that conformity pressure increases substantially with up to three majority members and stabilizes beyond that point \cite{forsythGroupDynamics2018,aschOpinionsSocialPressure1955,gerardConformityGroupSize1968,bondCultureConformityMetaanalysis1996}.

Together, these role-based manipulations enabled us to examine compliance behavior under controlled conditions without relying on participants’ pre-existing beliefs or personal characteristics. In this study, the terms 'majority' and 'minority' refer strictly to role-based positions created by numerical imbalance and assigned social power, rather than to any demographic attributes.

Based on these manipulations, participants were assigned to one of two roles:
\begin{itemize}
    \item \textbf{Majority with High Social Power (Three Seniors)}: Three participants per group received contextual information encouraging conservative and consensus-oriented decisions and were granted authority through the Senior role and reward allocation power.
    \item \textbf{Minority with Low Social Power (One Junior)}: One participant per group received contextual information encouraging a distinct perspective and occupied a lower-power role that depended on Senior evaluation.
\end{itemize}

Although all participants remained anonymous and were unaware of each other's personal information, they explicitly recognized their own role and the roles of other group members at the start of the study. These roles were announced before the discussion and were continuously visible in the chat interface, where each message was labeled as originating from a 'Senior' or a 'Junior'. All participants were also informed that an AI agent participated in the discussion and that system messages originated from the AI (\autoref{fig:systemImplementation}). In the AI-mediated Message condition, only the minority participant was informed about the private messaging feature. We made this design choice because the study focused on whether minority participants could experience psychological safety when expressing dissent through the system. To preserve this focus, we did not inform majority participants about the existence of the AI-mediated messaging feature.

By integrating both social power and majority influence mechanisms, we targeted compliance as the central experimental condition rather than examining each mechanism in isolation. This design allowed us to evaluate how effectively the system supported minority participants under realistic and consciously perceived compliance pressure.

\subsection{Participants}
We recruited 96 Korean participants (age $M$ = 26.60, $SD$ = 5.21, range = 19--42) and randomly assigned them into 24 groups of four. Each group consisted of three high-power majority members and one low-power minority member. Participants were recruited online and met the following inclusion criteria: Korean nationality, age 18 or older, prior experience in group decision-making, and familiarity with online chat environments. We assigned participants to roles using random assignment rather than matching based on individual characteristics. This approach allowed individual differences in communication style or argumentation tendency to distribute evenly across roles and experimental conditions. All participants completed the same tasks under an identical interface, time, and interaction constraints, which helped reduce systematic bias caused by individual-level variation. All participants received a briefing on the study procedures at the beginning of each session and were informed of their right to withdraw at any time. If any participant withdrew or did not provide consent, the session was canceled, and the remaining participants received 1,000 KRW as base compensation. All data were coded and de-identified to protect participant anonymity, including survey responses, interview transcripts, and chat records.

Demographic information included gender (61 female, 35 male) and education level: 46.9\% held a bachelor’s degree, 19.8\% a master’s degree, 15.6\% had some college education, 13.5\% completed high school or equivalent, and 4.2\% held doctoral degrees. On average, participants had 2.50 years ($SD$ = 3.15) of professional working experience. Additional background variables included self-reported familiarity with AI ($M$ = 4.83, $SD$ = 1.48), prior experience with group decision-making ($M$ = 5.01, $SD$ = 1.41), and online collaboration ($M$ = 4.39, $SD$ = 1.83). We determined the sample size at the condition level. Each experimental condition (Baseline, AIGC, AIMM) included at least 48 participants, which exceeded common minimum sample size conventions for group-based statistical analyses \cite{kwak2017central, hwangSoundSupportGendered2024}. Given the four-person group structure, this number represented the smallest feasible sample that maintained balanced group composition across conditions. Notably, 53.1\% of participants reported prior use of AI in group contexts. While this sample enabled consistent group composition and controlled role assignment, the relatively high AI familiarity and the Korean cultural context should be considered when interpreting generalizability. Details of background and demographic questionnaires are shown in \autoref{Appendix-demographic}.

\subsection{Task Description}
To create an immersive and compliance-inducing environment aligned with legitimate power roles, we designed two decision-making tasks that simulate realistic corporate scenarios. These tasks were selected to reflect the typical responsibilities and risk sensitivities associated with hierarchical roles in organizations, thereby reinforcing the assigned roles of Seniors (Majority with High Power) and Juniors (Minority with Low Power). The first task, team leader promotion, was adapted from prior studies in social psychology and AI-assisted group decision-making \cite{hwangSoundSupportGendered2024,binnsItsReducingHuman2018,laiScienceHumanAIDecision2021}. To ensure each participant experienced two distinct but structurally similar tasks, as required by the within-subject design, we developed a second task, the contractor selection task. This new task mirrors the decision logic of the first but is novel and was created specifically for this study.

Participants viewed task descriptions that differed by assigned role. This role-specific framing aimed to enhance task immersion by aligning each participant's goals with their social power (\autoref{Appendix-taskInstruction}):
\begin{itemize}
    \item Seniors were told they were responsible for the company’s long-term stability and reputation. They were instructed to make decisions that reflected conservative, proven judgment and risk mitigation.
    \item Juniors were told they needed to demonstrate their value to the organization by making bold, high-impact decisions. They were encouraged to pursue visible results even at the cost of taking risks. 
\end{itemize}

Each task presented three options: 1) A conservative option: low risk, long-term proven performance, but little immediate impact. 2) A challenging option: high potential but unproven, associated with visible outcomes and greater uncertainty. 3) A neutral option: a middle-ground choice that was relatively unattractive and intended as a control. Both tasks were intentionally structured to guide seniors toward the conservative option and juniors toward the challenging option, thereby creating natural opinion divergence between roles. This divergence was crucial to inducing majority-minority dynamics and enabling the study of compliance under controlled but immersive conditions.

\subsection{Experimental Procedure}
\begin{figure*}[]
  \centering
  \includegraphics[width=1.0\textwidth]{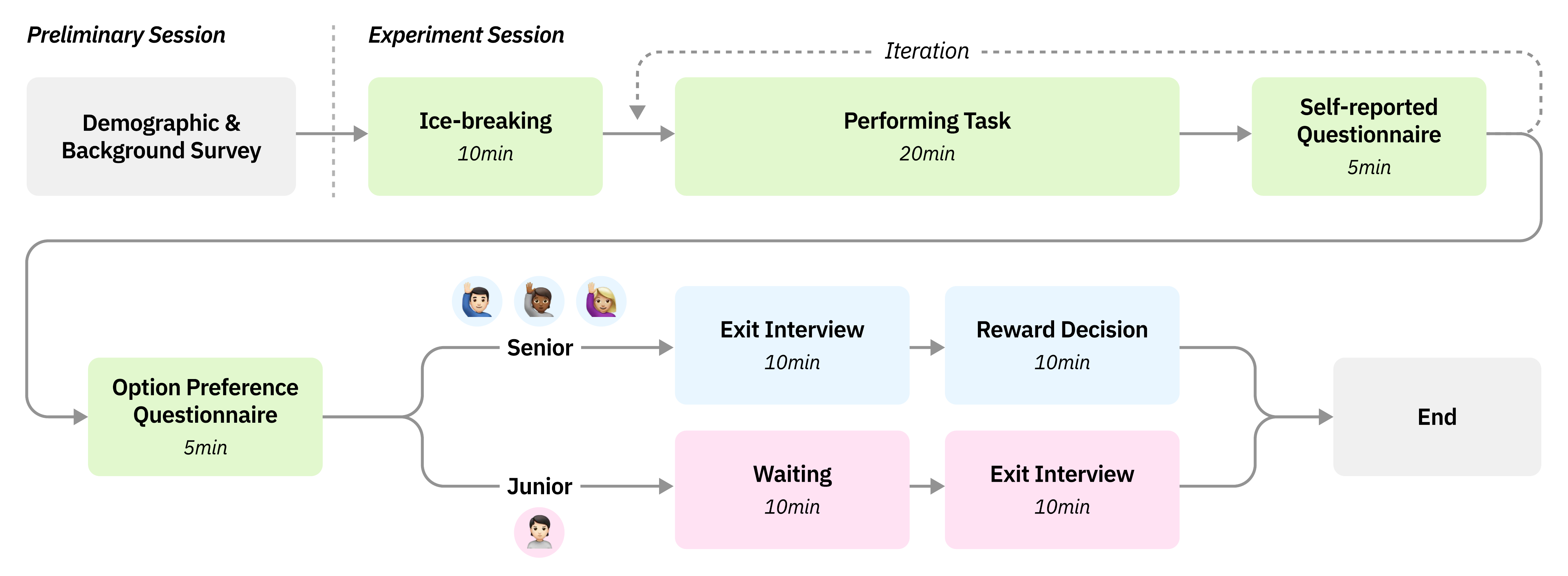}
  \caption{Overview of the experimental procedure: including pre-experiment surveys, ice-breaking, iterative decision-making tasks, post-task questionnaires, and role-specific exit interviews.}
  \Description{The experimental procedure includes a pre-experiment survey (5 min), ice-breaking (10 min), a decision-making task (25 min), and a post-task questionnaire (5 min). Participants then complete an option preference survey (5 min) followed by role-specific activities: seniors conduct an exit interview (10 min) and make a reward decision (10 min), while juniors wait (10 min) before their exit interview (10 min).}
  \label{fig:procedure}
\end{figure*}

Each session lasted approximately 90 to 105 minutes and was organized into three main phases (\autoref{fig:procedure}): pre-experiment setup, main task session, and post-experiment survey and interviews. The procedure was designed to ensure immersion and maintain power dynamics.

Before the main tasks, participants completed a series of preparatory activities online on the day of the experiment:
\begin{itemize}
    \item \textbf{\textit{Demographic and Background Survey}}: Participants submitted information on age, education, work experience, and prior familiarity with AI, group decision-making, and online collaboration.
    \item \textbf{\textit{Agreement to Participation}}: Participants reviewed consent materials and confirmed their participation. If any participant declined, the session was canceled, and the remaining participants received 1,000 KRW compensation.
    \item \textbf{\textit{Ice-breaking Activity (10 min)}}: Using an anonymous commercial chatting platform such as \footnote{KakaoTalk: https://www.kakaocorp.com/page/service/service/KakaoTalk?lang=en}{KakaoTalk}, participants introduced themselves using pseudonyms, created a team name, and collaboratively defined a slogan to become familiar with anonymous online chat environments. This activity aimed to establish group cohesion while maintaining role-based power distinctions.
\end{itemize}

Each group completed two decision-making blocks, with the system condition (Baseline + either AIGC or AIMM) and task order counterbalanced across sessions:
\begin{itemize}
    \item \textbf{\textit{Decision-Making Task (20 min)}}: Participants engaged in structured group discussions within an experimental chat platform (\autoref{fig:systemImplementation}). Each task presented the three options designed to create opinion divergence between Seniors and juniors.
    \item \textbf{\textit{Self-Reported Questionnaire (5 min)}}: After each task, participants completed a survey assessing psychological safety, decision satisfaction, cognitive load, and perception of interaction with each devil's advocate.
\end{itemize}

The session concluded with role-specific exit procedures:
\begin{itemize}
    \item \textbf{\textit{Option Preference Questionnaire (5 min)}}: Each participant indicated how strongly they preferred their chosen option for their assigned role in each task, reflecting how immersed they felt in the given situation.
    \item \textbf{\textit{Senior Exit Interview \& Reward Decision (20 min)}}: The three seniors participated in a joint \footnote{Zoom: https://www.zoom.com/}{Zoom} interview, reflecting on the group’s performance and dynamics. Then, the seniors jointly decided whether to allocate a bonus reward to the Junior, reinforcing the reward-based power structure.
    \item \textbf{\textit{Junior Exit Interview (10 min)}}: After a brief waiting period during the reward decision phase, the Junior completed a private interview to share their individual experiences.
\end{itemize}

Although all participants received equal final compensation (20,000 KRW), the process preserved the perception of differential power, crucial for studying compliance and communication behavior.

\subsection{Measurement}
This study employed both self-reported and objective measures to assess how system conditions and participant type influenced group dynamics. Self-reported measures captured participants’ subjective experiences using 7-point Likert scales (1 = strongly disagree, 7 = strongly agree). These included the agreement questionnaire, psychological safety, satisfaction with the decision-making process and outcome, cognitive workload, and perceptions of the AI system. Objective behavioral measures were used to quantify participants' engagement in the group discussion (\autoref{Appendix-selfReported}).

\begin{itemize}
    \item \textbf{\textit{Validation of Induced Opinion (Study Premise)}}: To confirm that the experimental tasks successfully induced role-based opinion divergence, participants rated their preference for each of the three decision options after the main experiment. Ratings were collected using a 7-point Likert scale. Participants were expected to prefer the option that matched their assigned role in each situation, with seniors choosing the conservative option (option 1) and juniors choosing the ambitious option (option 2). This served as a manipulation check to validate the foundation of our compliance-oriented design.
    \item \textbf{\textit{Psychological Safety and Marginalization (RQ1)}}: We assessed participants’ feelings of safety in expressing dissent using established measures of psychological safety and marginalization. These items gauged the extent to which participants felt heard, supported, and free to express disagreement within their group \cite{edmondsonPsychologicalSafetyLearning1999, castilloConstructionValidationIntragroup2007, hwangSoundSupportGendered2024, janisVictimsGroupthinkPsychological1972}.
    \item \textbf{\textit{Engagement in Group Discussion (RQ2)}}: Engagement was the only objective behavioral metric used in this study. It was operationalized as each participant’s level of contribution to the conversation, measured by 1) the number of messages sent and 2) the total number of characters typed during the task.
    \item \textbf{\textit{Perception of Decision-Making Process and Outcomes (RQ3)}}: Participants rated the quality of the group decision-making process across several dimensions, including influence, group cohesion, support from teammates, and consideration of diverse opinions \cite{chiangEnhancingAIAssistedGroup2024, ganoticeTeamCohesivenessCollective2022, liImprovingNonNativeSpeakers2022, cookeMeasuringTeamKnowledge2000, easleyRelatingCollaborativeTechnology2003}. Decision outcome quality was assessed through satisfaction and perceived validity of the group’s final choice \cite{chenUserSatisfactionGroup2012, paulUserSatisfactionSystem2004, carneiroPredictingSatisfactionPerceived2019, lopesValidationGroupDecisions2014, woodParticipationInfluenceSatisfaction1972}.
    \item \textbf{\textit{Cognitive Workload (RQ4)}}: Cognitive workload was measured using the NASA Task Load Index (NASA-TLX), which assesses task difficulty across five dimensions, including mental demand, temporal demand, performance, effort, and frustration \cite{hartDevelopmentNASATLXTask1988}.
    \item \textbf{\textit{Perception of the AI System (Exploratory)}}: To contextualize how participants experienced AI-mediated support, we collected ratings across four dimensions: cooperation, satisfaction, quality, and fairness \cite{chiangEnhancingAIAssistedGroup2024, reinkemeierCanHumanizingVoice2022, yuanWordcraftStoryWriting2022}. These measures assessed user trust and acceptance of the AI’s role in shaping group dynamics.
\end{itemize}

Data were analyzed using robust linear mixed models with random effects, suitable for the repeated-measures design and small-group variance. Bonferroni post-hoc tests compared outcomes across experimental conditions and participant types.

In addition to quantitative and behavioral measures, semi-structured exit interviews captured participants’ subjective experiences. Juniors participated in one-on-one interviews, while Seniors took part in a group interview. Key topics included role immersion, psychological safety, group dynamics, and perceptions of the AI system. For the AIMM condition, juniors were also asked about their experience with the secret messaging feature. Exit interview questions addressed comfort in expressing opinions, experiences of conformity or pressure, and the perceived impact of AI during discussions. After obtaining consent, all interviews were recorded and transcribed using a commercial speech-to-text service (\footnote{Clova Note: clovanote.naver.com}{Clova Note}). Interview transcripts were briefly reviewed to identify common themes that could help explain the quantitative results.

\section{Findings}
The experimental results showed that the senior and junior participants had different decision-making patterns. Juniors preferred challenging options, while seniors favored stable ones, with final group decisions aligning with senior preferences about 80\% of the time. LLM-powered minority support had mixed impacts: AI counterarguments somewhat improved junior satisfaction, but AI-mediated communication increased their cognitive load. While seniors' experiences remained stable across conditions, juniors' psychological safety and satisfaction varied based on the AI interventions.

Across all analyses, we examined the effects of system Condition, participant Role, and their interaction using regression models tailored to the data structure of each outcome. Given the repeated-measures design, group-level dependencies, and unequal cell sizes inherent in role-based group experiments, we adopted robust estimation to obtain stable coefficient estimates under realistic violations of normality and homoscedasticity. For outcomes involving repeated observations (including self-reported measures, validation of majority--minority manipulation, and engagement metrics), we employed robust linear mixed-effects models with participant-level random intercepts (estimated using \texttt{rlmer} from the \texttt{robustlmm} package in R), with \textbf{Baseline} and \textbf{Junior} as reference levels:

\[
Y_{ij} = \beta_0 + \beta_1 \textit{Condition}_{ij} + \beta_2 \textit{Role}_{i} + \beta_3 (\textit{Condition} \times \textit{Role})_{ij} + u_i + \epsilon_{ij},
\]

\noindent where $i$ indexes participants, $j$ indexes tasks, $\beta_0$ is the intercept, $\beta_1$--$\beta_3$ are regression coefficients, $u_i$ is the participant-level random intercept, and $\epsilon_{ij}$ is the residual. For perception-of-agent measures, which were collected once per participant within each AI-assisted condition (AIGC or AIMM) without repeated observations, we used \textbf{robust linear regression} (estimated using \texttt{rlm} from the \texttt{MASS} package in R), with \textbf{AIGC} and \textbf{Junior} as reference levels:

\[
Y_i = \beta_0 + \beta_1 \textit{Condition}_i + \beta_2 \textit{Role}_i + \beta_3 (\textit{Condition} \times \textit{Role})_i + \epsilon_i,
\]

\noindent where $i$ indexes participants, $\beta_0$ is the intercept, $\beta_1$--$\beta_3$ are regression coefficients, and $\epsilon_i$ is the residual.

The following sections examine role-based preferences, the effects of AI interventions on psychological safety (RQ1), engagement patterns (RQ2), decision-making experience and satisfaction (RQ3), cognitive workload (RQ4), and emergent ethical implications. All measured details are in \autoref{appendix-detailedMeasure}.

\begin{figure*}[]
  \centering
  \includegraphics[width=1.0\textwidth]{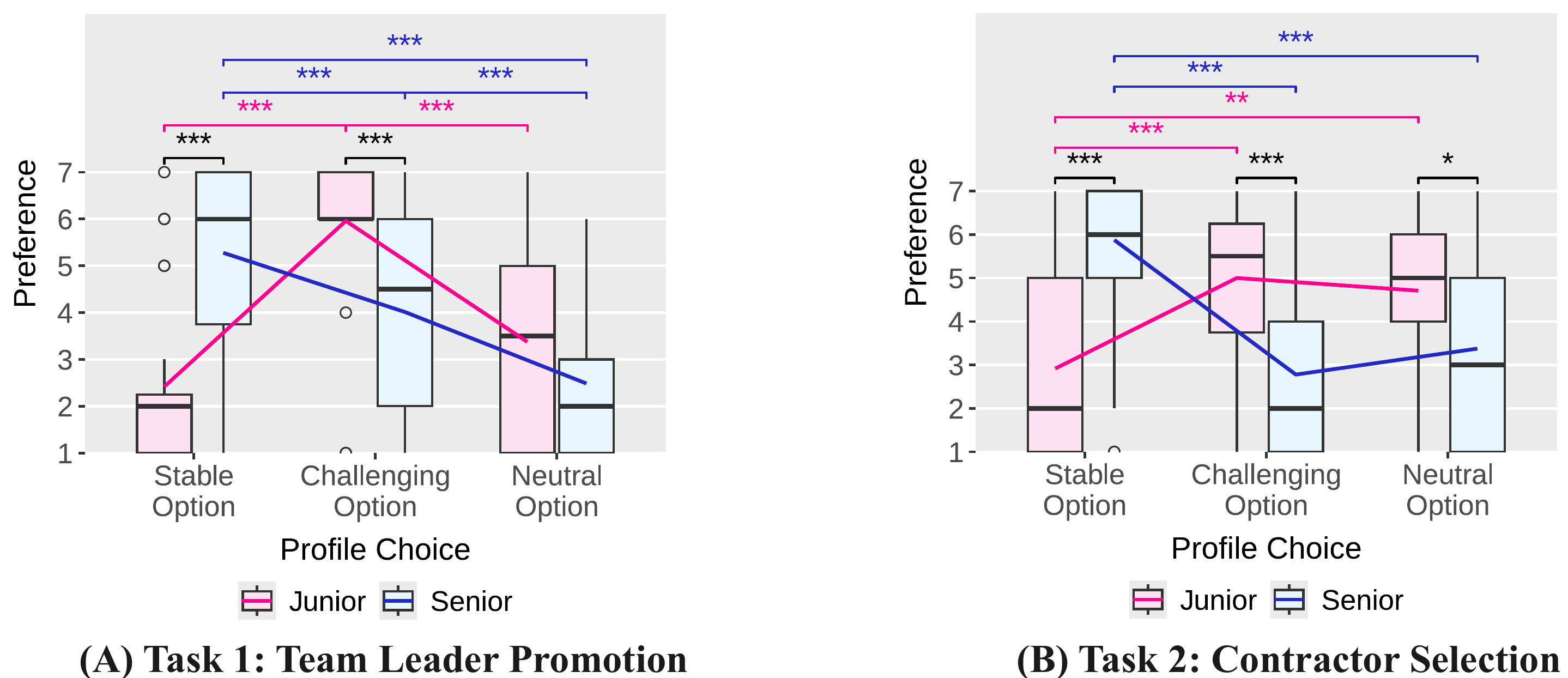}
  \caption{Role-based differences in option preferences for (A) Task 1 and (B) Task 2. Preferences were measured on a 7-point Likert scale, with seniors favoring stable options (Option 1), while juniors preferred challenging alternatives (Option 2). Neutral options (Option 3) were generally rated lower by both roles, reflecting distinct preference patterns driven by role dynamics. Brackets indicate statistically significant pairwise differences based on Bonferroni-adjusted post-hoc tests. Only significant comparisons are shown. (*$p < .05$; **$p < .01$; ***$p < .001$)}
  \Description{The boxplots illustrate role-based preference differences on a 7-point Likert scale for Task 1 (A) and Task 2 (B). Seniors consistently favored stable options (Option 1), while juniors showed a stronger preference for challenging alternatives (Option 2). Neutral options (Option 3) were rated lower across roles, emphasizing how role dynamics influence decision preferences.}
  \label{fig:selfReportedImmersion}
\end{figure*}

\begin{table*}[t]
\centering
\caption{Robust regression coefficients ($\beta$) and standard errors (SE) for the \textit{Validation of Majority \& Minority Manipulation}.  
Baseline is \textit{Option 1 – Junior}. Stars denote significance (* $p<.05$, ** $p<.01$, *** $p<.001$).}
\small
\resizebox{\textwidth}{!}{%
\begin{tabular}{lcccccc}
\toprule
 & \multicolumn{6}{c}{Predictors} \\
\cmidrule(lr){2-7}
Task & Intercept & Option 2 vs.\ Option 1 & Option 3 vs.\ Option 1 & Senior vs.\ Junior & Option 2$\times$Senior & Option 3$\times$Senior \\
\midrule
Task 1 & 2.26 (0.42)*** & 3.89 (0.59)*** & 1.08 (0.59) & 3.29 (0.48)*** & -5.42 (0.68)*** & -4.22 (0.68)*** \\
Task 2 & 2.76 (0.37)*** & 2.38 (0.52)*** & 2.02 (0.52)*** & 3.28 (0.43)*** & -5.70 (0.61)*** & -4.73 (0.61)*** \\
\bottomrule
\end{tabular}}
\label{tab:manip_validation_rlm}
\end{table*}

\subsection{Validation of Majority and Minority Manipulation (Experimental Setup)}
To validate the experimental setup, we examined whether the role-based preference manipulation effectively created consistent majority and minority positions. Each task presented participants with three options: a stable but less innovative choice (Option 1), a more challenging alternative (Option 2), and a neutral option (Option 3). Seniors were expected to prefer the stability of Option 1, while juniors were guided toward the more ambitious Option 2. As shown in \autoref{tab:manip_validation_rlm} and \autoref{fig:selfReportedImmersion}, participants’ choices aligned with this design: seniors consistently rated Option 1 highest ($M$ = 5.28, $SD$ = 2.21 for task 1, $M$ = 5.88, $SD$ = 1.58 for task 2), and juniors rated Option 2 highest ($M$ = 5.96, $SD$ = 1.68 for task 1, $M$ = 5.00, $SD$ = 1.89 for task 2). These divergent patterns emerged clearly across both tasks, with statistically significant differences between roles for both primary options. The neutral Option 3 was consistently rated lower, suggesting it did not attract strong preference from either group. These results confirm that the role framing induced the intended preference structures, effectively producing conditions where one viewpoint dominated in each group while another remained in the minority. Notably, participants showed slightly more openness to challenging alternatives in the Team Leader Promotion task and favored more stable choices in the Contractor Selection task, reflecting task-specific variation within the overall successful manipulation.

\subsection{Psychological Safety (RQ1)}

\begin{figure*}[]
  \centering
  \includegraphics[width=1.0\textwidth]{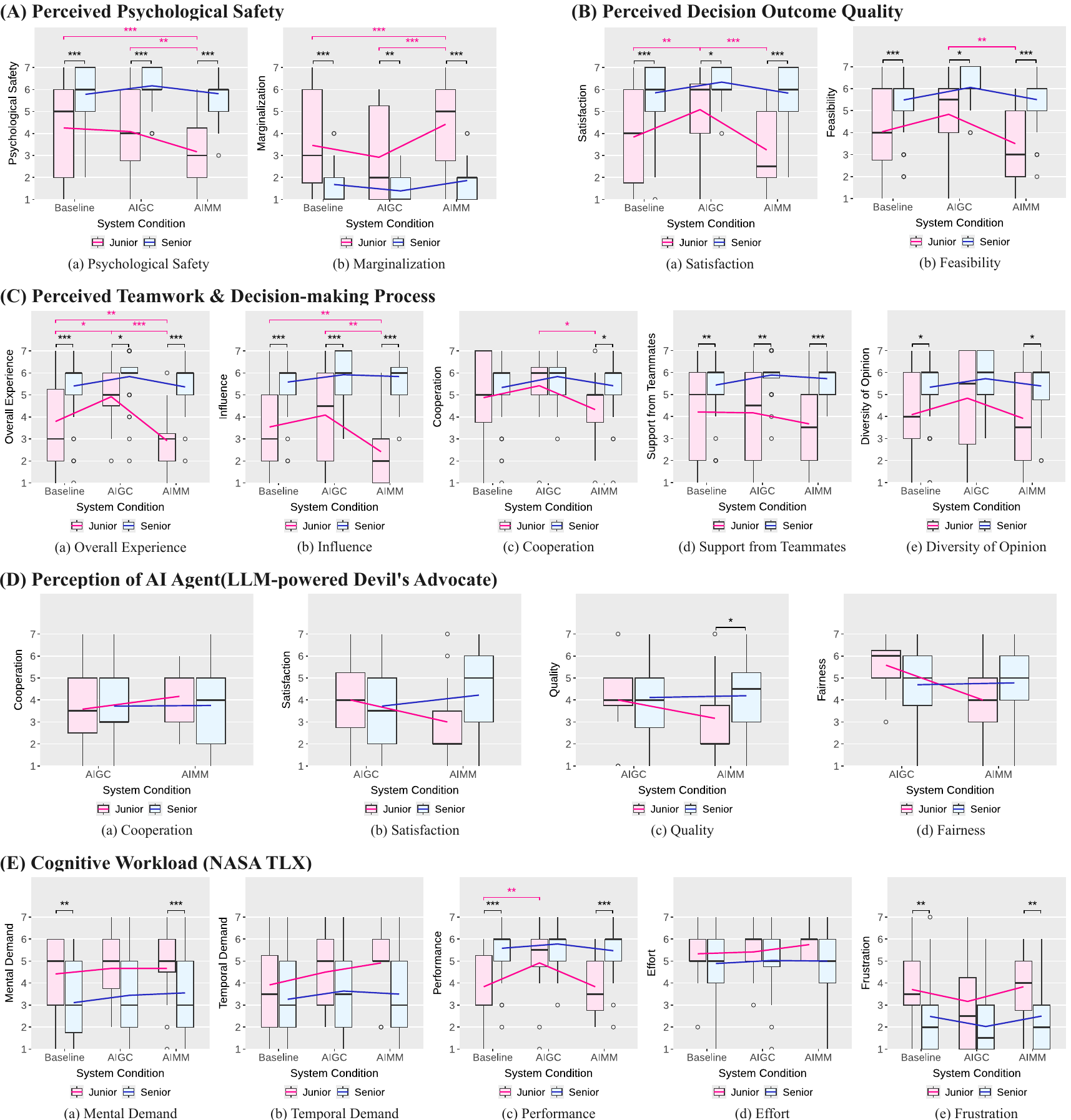}
  \caption{Self-reported metrics across conditions (Baseline, AIGC Condition, AIMM Condition) for psychological safety, decision outcome quality, teamwork, workload (NASA-TLX), and perceptions of the LLM-powered Devil's Advocate. Each subfigure compares Junior and Senior participants’ responses under each condition. Pink and blue lines represent Junior and Senior participants, respectively. Asterisks and brackets indicate statistically significant pairwise differences based on Bonferroni-adjusted post-hoc comparisons. Only comparisons with statistically significant differences are shown. (*$p < .05$; **$p < .01$; ***$p < .001$)}
  \Description{The figure compares self-reported metrics across conditions (A, B, C). Juniors report higher psychological safety, decision satisfaction, and fairness in Condition C, while teamwork and workload metrics show moderate improvements. The AI (Devil's Advocate) is rated more favorably in Condition C, particularly by juniors, highlighting its role in fostering inclusion and decision quality.}
  \label{fig:selfReported}
\end{figure*}

Quantitative results show that the manipulation of AI roles shaped participants’ perceptions of psychological safety and marginalization, particularly among juniors. As shown in \autoref{tab:robust_coeffs_selfReported}-(A) and \autoref{fig:selfReported}-(A), juniors reported the lowest psychological safety in the AIMM condition ($M$=3.17, $SD$=1.53), compared to the Baseline ($M$=4.25, $SD$=2.05) and AIGC conditions ($M$=4.08, $SD$=2.15). A robust linear mixed model confirmed a significant drop in psychological safety for juniors in AIMM relative to Baseline ($\beta$=-1.40, $SE$=0.28, $p$<.001), as well as a significant interaction effect with role ($\beta$=1.49, $SE$=0.32, $p$<.001). In contrast, seniors reported consistently high psychological safety across all conditions, with no meaningful variation.

Marginalization scores followed a similar pattern. Juniors in AIMM reported the highest levels of marginalization ($M$=4.42, $SD$=2.02), compared to Baseline ($M$=3.46, $SD$=2.23) and AIGC ($M$=2.92, $SD$=2.19). Regression results indicated a significant increase in marginalization in AIMM ($\beta$=0.96, $SE$=0.22, $p$<.001), along with a strong role-by-condition interaction effect ($\beta$=-0.88, $SE$=0.25, $p$<.001). Seniors, by comparison, consistently reported low marginalization across all three systems.

Qualitative interviews help explain this mismatch between design intent and user experience. Juniors initially believed that having the AI express their views would help them be heard. However, many reported that their AI-mediated messages were dismissed or overlooked. As one junior shared,
\begin{quote}
    \textit{I thought that by the AI putting forward my opinion, my opinion would be more recognized, but that was not the case, so I was a little intimidated.} (P96)
\end{quote}

Seniors also acknowledged disregarding the AI’s input:
\begin{quote}
    \textit{It's an AI, so I just kind of ignored it.} (P6)
\end{quote}
\begin{quote}
    \textit{The fact that it wasn't a person made the AI's words carry less weight.} (P71)
\end{quote}

These accounts help explain why juniors reported the lowest psychological safety and highest marginalization in the AIMM condition: while the system was designed to protect minority voices, it inadvertently removed speaker agency and visibility. By contrast, the AIGC condition, where the AI presented generalized counterarguments, was more effective in reducing marginalization without evoking the same social discounting. These findings suggest that anonymity mechanisms, though well-intentioned, can sometimes backfire if they obscure the source of dissenting input.

\begin{figure*}[]
  \centering
  \includegraphics[width=0.95\textwidth]{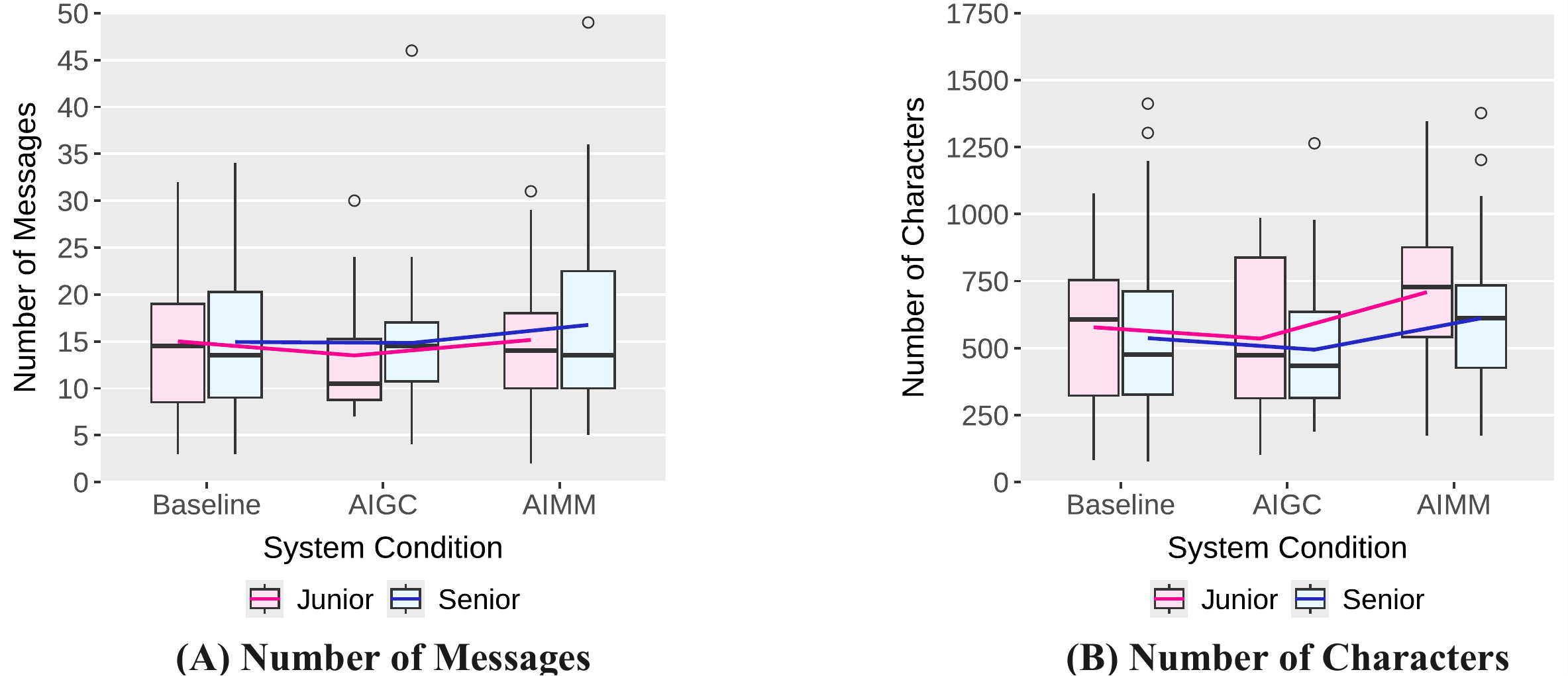}
  \caption{Contribution and engagement patterns across conditions (Baseline, AIGC, AIMM) measured by (A) number of messages, (B) number of characters typed. No significant differences were found in Bonferroni-corrected post-hoc tests.}
  \Description{The boxplots illustrate contribution and engagement patterns across conditions (Baseline, AIGC, AIMM) for seniors (green) and juniors (orange). Metrics include (A) the number of messages, showing similar distribution across roles, (B) the number of characters typed, with juniors generally contributing more text, and (C) normalized engagement scores, reflecting comparable engagement across roles and conditions. These measures highlight interaction dynamics and role-based contributions.}
  \label{fig:dialogueVis}
\end{figure*}

\begin{table*}[t]
\centering
\caption{Robust regression coefficients ($\beta$) and standard errors (SE) for communication volume.  
Baseline is \textit{Baseline – Junior}. Stars denote significance (* $p<.05$, ** $p<.01$, *** $p<.001$).}
\small
\resizebox{\textwidth}{!}{%
\begin{tabular}{lcccccc}
\toprule
 & \multicolumn{6}{c}{Predictors} \\
\cmidrule(lr){2-7}
Outcome & Intercept & AIGC vs.\ Baseline & AIMM vs.\ Baseline & Senior vs.\ Junior & AIGC$\times$Senior & AIMM$\times$Senior \\
\midrule
Number of Messages & 13.78 (1.52)*** & -0.05 (1.76) & 0.05 (1.74) & 0.51 (1.76) & 0.19 (2.03) & 0.98 (2.01) \\
Number of Characters & 558.46 (59.80)*** & -27.73 (61.93) & 129.95 (61.29)* & -45.13 (69.23) & -4.08 (71.53) & -25.50 (70.98) \\
\bottomrule
\end{tabular}}
\label{tab:comm_volume_rlmer}
\end{table*}

\subsection{Engagement in Group Discussion (RQ2)}

We analyzed participant engagement using two metrics: the number of messages and the number of characters typed. As shown in \autoref{tab:comm_volume_rlmer} and \autoref{fig:dialogueVis}, the number of messages did not significantly differ by condition or role, suggesting stable turn-taking patterns across all settings (e.g., juniors in Baseline: $M$=15.00, $SD$=8.03). 

However, the AIMM condition was associated with an increase in the number of characters typed. For example, juniors in AIMM typed more characters ($M$=708.62, $SD$=319.58) than in the Baseline ($M$=577.62, $SD$=279.56) and AIGC conditions ($M$=535.42, $SD$=301.04). This increase was significant in the regression ($\beta$=129.95, $p$<.05), though not after Bonferroni correction (\autoref{fig:dialogueVis}), suggesting that the AI agent prompted juniors to elaborate more, even as message counts held steady.

This trend is supported by post-task interviews. Participants in AIMM noted that the AI agent helped amplify juniors’ contributions, encouraging them to participate more actively. As one senior with AIMM explained,
\begin{quote}
    \textit{I feel like at least one person is on the junior's side, so I think a junior is a little more willing to give his opinion.} (P59)
\end{quote}

These observations suggest that while the AIMM condition did not change the number of turns participants took, it may have subtly increased the depth or length of their contributions. This implies that the AIMM condition may have had a motivating effect, prompting participants to elaborate more even if their relative share of discussion remained unchanged.

\begin{table*}[t]
\centering
\caption{Robust regression coefficients ($\beta$) and standard errors (SE) for each self‑report measure. Baseline is \textit{Baseline – Junior}. Stars denote significance (* $p<.05$, ** $p<.01$, *** $p<.001$).}
\small
\resizebox{\textwidth}{!}{%
\begin{tabular}{lcccccc}
\toprule
 & \multicolumn{6}{c}{Predictors} \\
\cmidrule(lr){2-7}
Measure & Intercept (Baseline, Junior) & AIGC vs.\ Baseline & AIMM vs.\ Baseline & Senior vs.\ Junior & AIGC$\times$Senior & AIMM$\times$Senior \\
\midrule
\multicolumn{7}{l}{\textbf{(A) Perceived Psychological Safety}}\\
Psychological Safety & 4.57 (0.23)*** & -0.01 (0.28) & -1.40 (0.28)*** & 1.25 (0.27)*** & 0.29 (0.32) & 1.49 (0.32)*** \\
Marginalization      & 2.99 (0.20)*** & -0.53 (0.22)* &  0.96 (0.22)*** & -1.32 (0.22)*** & 0.30 (0.25) & -0.88 (0.25)*** \\
\addlinespace
\multicolumn{7}{l}{\textbf{(B) Perceived Decision Outcome Quality}}\\
Satisfaction         & 4.05 (0.23)*** &  1.17 (0.31)*** & -0.63 (0.31)*  & 1.89 (0.26)*** & -0.76 (0.35)* & 0.71 (0.35)* \\
Feasibility          & 4.31 (0.24)*** &  0.69 (0.29)*  & -0.67 (0.29)*  & 1.30 (0.27)*** & -0.27 (0.33) & 0.77 (0.33)* \\
\addlinespace
\multicolumn{7}{l}{\textbf{(C) Perceived Teamwork \& Decision‑making Process}}\\
Overall Experience   & 3.85 (0.23)*** &  0.89 (0.30)** & -1.08 (0.30)*** & 1.68 (0.26)*** & -0.59 (0.34) & 1.09 (0.35)** \\
Influence            & 3.50 (0.24)*** &  0.43 (0.33)   & -1.24 (0.33)*** & 2.16 (0.28)*** & -0.19 (0.38) & 1.50 (0.38)*** \\
Cooperation          & 5.22 (0.25)*** &  0.64 (0.34)   & -0.85 (0.35)** & 0.22 (0.28)    & -0.21 (0.40) & 0.94 (0.40)*  \\
Support from Teammates & 4.49 (0.24)*** & 0.01 (0.36)   & -1.04 (0.36)** & 1.05 (0.28)*** & 0.38 (0.41) & 1.26 (0.41)** \\
Diversity of Opinion & 4.18 (0.33)*** &  1.03 (0.56)   & -0.31 (0.56)   & 1.25 (0.38)*** & -0.68 (0.65) & 0.37 (0.65)   \\
\addlinespace
\multicolumn{7}{l}{\textbf{(D) Cognitive Workload (NASA‑TLX)}}\\
Mental Demand        & 4.68 (0.37)*** & -0.24 (0.38)   &  0.81 (0.38)*  & -1.61 (0.43)*** & 0.57 (0.44) & -0.53 (0.44) \\
Temporal Demand      & 3.92 (0.43)*** &  0.40 (0.52)   &  1.08 (0.52)*  & -0.66 (0.50)    & -0.08 (0.60) & -0.78 (0.60) \\
Performance          & 4.01 (0.23)*** &  1.13 (0.31)*** & 0.12 (0.31)   & 1.61 (0.26)*** & -0.97 (0.35)** & -0.15 (0.35) \\
Effort               & 5.40 (0.27)*** &  0.08 (0.46)   &  0.35 (0.46)   & -0.39 (0.31)    & 0.06 (0.53) & -0.15 (0.53) \\
Frustration          & 3.59 (0.31)*** & -0.57 (0.38)   &  0.40 (0.38)   & -1.23 (0.35)*** & 0.24 (0.44) & -0.46 (0.44) \\
\bottomrule
\end{tabular}}
\label{tab:robust_coeffs_selfReported}
\end{table*}

\begin{table*}[t]
\centering
\caption{Robust regression coefficients ($\beta$) and standard errors (SE) for self‑reported \textit{Perception of AI}.  
Reference level = \textit{AIGC – Junior}. Stars denote significance (* $p<.05$, ** $p<.01$, *** $p<.001$).}
\small
\resizebox{\textwidth}{!}{%
\begin{tabular}{lcccc}
\toprule
 & \multicolumn{4}{c}{Predictors} \\
\cmidrule(lr){2-5}
Measure & Intercept (AIGC, Junior) & AIMM vs.\ AIGC & Senior vs.\ Junior & AIMM$\times$Senior \\
\midrule
Cooperation  & 3.51 (0.51)*** & 0.66 (0.72)  & 0.18 (0.59) & -0.62 (0.83) \\
Satisfaction & 4.00 (0.55)*** & -1.19 (0.78) & -0.29 (0.64) & 1.74 (0.91) \\
Quality      & 4.14 (0.55)*** & -1.46 (0.78) & 0.04 (0.64) & 1.66 (0.90) \\
Fairness     & 5.61 (0.46)*** & -1.61 (0.66)* & -0.79 (0.54) & 1.64 (0.76)* \\
\bottomrule
\end{tabular}}
\label{tab:ai_perception_rlm}
\end{table*}

\subsection{Satisfaction with Decision-making Processes and Outcomes (RQ3)}
\subsubsection{Satisfaction with Decision-making Process}

Perceptions of the decision-making process revealed a clear divide between juniors and seniors, particularly in the AIMM condition. While seniors' ratings remained relatively stable across all conditions (e.g., overall experience $M$=5.36, $SD$=1.02 in AIMM), juniors experienced a sharp decline in satisfaction, influence, cooperation, and support from teammates when interacting through the AI-mediated message relay. For instance, juniors’ overall satisfaction dropped to $M$=2.92 ($SD$=1.51) in AIMM, compared to $M$=3.79 in Baseline and $M$=4.92 in AIGC. A similar trend emerged in perceived influence ($M$=2.42 in AIMM) and cooperation ($M$=4.33), showing clear declines across these measures (see \autoref{tab:robust_coeffs_selfReported}-(C), \autoref{fig:selfReported}-(C)).

Among the three conditions, juniors reported generally more positive experiences in the AIGC condition, where the AI provided generalized counterarguments. In contrast, the AIMM condition—designed to support dissent by anonymously relaying minority opinions—appeared to backfire. Juniors reported feeling excluded, with limited influence over the discussion and little support from teammates.

These declines were not observed among seniors, who consistently rated the decision-making process positively, regardless of condition. Interaction effects in the regression models confirmed that these role-based gaps were statistically significant across most measures, particularly for satisfaction, influence, and cooperation.

Interview data help explain these patterns. Several juniors shared that, despite expecting the AI to help convey their views more safely, they felt ignored when their ideas were voiced anonymously. As one participant put it,
\begin{quote}
    \textit{It wasn't just me that had a different opinion, but the devil agent was now giving a little bit of a dissenting opinion, so I felt like I wasn't the only one who stood out from the group.} (P76)
\end{quote}

Others described how the AIGC condition helped create a more fluid and open atmosphere:
\begin{quote}
    \textit{The AI kept arguing back rather than directly helping, which made the atmosphere more fluid and made me see the seniors' point of view again.} (P20)
\end{quote}

Seniors also acknowledged the benefits of the AIGC agent in encouraging broader consideration:
\begin{quote}
    \textit{When AI said we should consider another option, I felt like that was a positive direction.} (P78)
\end{quote}

Together, these findings suggest that while both systems aimed to promote more inclusive decision-making, the AIGC approach better supported junior participation without compromising the overall group dynamic.

\subsubsection{Satisfaction with Decision-making Outcome}

Perceived decision outcome quality differed significantly between roles and across conditions (\autoref{tab:robust_coeffs_selfReported}-(B) \& \autoref{fig:selfReported}-(B)). Seniors consistently reported high satisfaction and feasibility of outcomes, with only modest variation across conditions. In contrast, juniors responded more sensitively to the system design. Their satisfaction and feasibility ratings improved in the AIGC condition (e.g., satisfaction $M$=5.08, $SD$=1.98), but dropped sharply in the AIMM condition ($M$=3.25, $SD$=1.76), widening the gap between the two roles. This pattern also held for perceived feasibility (AIMM: $M$=3.50, AIGC: $M$=4.83), with juniors in AIMM reporting the lowest ratings across all conditions.

The AIMM condition, intended to support minority perspectives, appears to have backfired from the perspective of outcome satisfaction. While seniors' scores remained high across all conditions, juniors expressed frustration that the AI-mediated messaging did not meaningfully affect final decisions, which continued to reflect the majority’s view. As one participant noted,
\begin{quote}
    \textit{If the outcome is the same... it's better to just make the decision without the AI, because I don't think it changes the psychological pressure that the juniors feel or the seniors' opinions.} (P92)
\end{quote}

This sentiment reflects a broader concern: when dissenting input is filtered through a system that lacks perceived influence, users may become disengaged from the outcome. In contrast, the AIGC condition modestly improved perceived outcome quality among both juniors and seniors by creating space for alternative perspectives, even if the final decisions remained largely unchanged.

\subsubsection{Perception of LLM-powered Devil's Advocate}

Participants’ perceptions of the AI agent varied modestly across conditions, with some emerging differences between juniors and seniors. Overall, seniors’ ratings remained relatively steady, while juniors showed a slight decline in satisfaction, perceived quality, and fairness in the AIMM condition, where the AI anonymously relayed the minority members' dissenting opinions. For instance, juniors in AIMM rated their satisfaction with the AI at $M$=3.00 ($SD$=1.95), compared to seniors at $M$=4.22 ($SD$=1.79), a statistically significant difference (post-hoc contrast = -1.45, $p$=0.023) (\autoref{tab:ai_perception_rlm} \& \autoref{fig:selfReported}-(D)).

Cooperation was perceived similarly across roles and conditions (e.g., juniors in AIMM: $M$=4.17), indicating that participants generally accepted the AI’s involvement in the discussion process. However, juniors reported somewhat lower fairness in AIMM ($M$=4.00) compared to AIGC ($M$=5.58), with this difference reaching significance. A similar trend appeared for perceived quality ($M$=3.17 in AIMM vs.\ $M$=4.00 in AIGC), with juniors again rating the agent lower than seniors in AIMM.

These trends, while not always robust across all measures, suggest that the style of mediation may have shaped juniors’ impressions of the AI in subtle but consequential ways. 

Qualitative feedback helps contextualize these responses. As discussed earlier, some juniors felt that the AI-mediated contributions were overlooked or failed to influence the group dynamic. This perceived lack of impact may have tempered their views of the AI’s helpfulness. In contrast, seniors appeared less affected by how the AI was implemented, maintaining neutral-to-positive perceptions regardless of condition. Taken together, these findings suggest that while the AI agent was generally accepted, its perceived effectiveness—especially in the AIMM condition—was more sensitive for the group it was designed to support.

\subsection{Cognitive Workload (RQ4)}

Cognitive workload ratings showed modest but consistent differences between roles, with juniors generally reporting higher mental and temporal demands, greater frustration, and lower perceived performance than seniors. While most differences were not statistically significant, juniors in the AIMM condition reported the highest cognitive load across several dimensions, including mental demand ($M$=4.67, $SD$=1.78), temporal demand ($M$=4.92, $SD$=1.51), and frustration ($M$=3.83, $SD$=1.70), as shown in \autoref{tab:robust_coeffs_selfReported}-(D) and \autoref{fig:selfReported}-(E).

These patterns suggest that although the AIMM condition did not significantly elevate workload scores overall, it introduced added complexity for juniors. Notably, within the AIMM condition, juniors reported significantly higher temporal demand than seniors (post-hoc contrast=1.44, $SE$=0.62, $p$=0.020), and performance satisfaction—which improved in the AIGC condition ($M$=4.92, $SD$=1.78)—returned to baseline levels in AIMM ($M$=3.83, $SD$=1.53), suggesting that the benefits of AI support were not sustained under the anonymous messaging setup.

In contrast, seniors’ ratings remained relatively stable across conditions. Perceived effort was comparable across all roles and systems, with no significant differences, indicating that all participants felt they were trying equally hard regardless of the system design.

Interview data help contextualize juniors’ elevated workload in AIMM. Several participants described cognitive strain from juggling task comprehension, opinion formulation, and coordination with the AI agent. One junior explained,
\begin{quote}
\textit{Because I have to look at the task material and understand the situation... I have to decide what to say to the AI and what opinion I will give... I think it was hard because I had so many things to think about during that time...} (P8)
\end{quote}

Others mentioned the delayed timing of AI responses as a source of additional burden:
\begin{quote}
\textit{It was kind of hard to get my opinion across right away and at the right time because you have to wait eight turns for the devil agent to speak.} (P60)
\end{quote}

Participants also expressed uncertainty about how to manage their input through the AI:
\begin{quote}
\textit{First of all, when to turn it off, that was the most questionable thing for me, so it was hard for me to say when to turn it off and when to say my opinion.} (P52)
\end{quote}

Together, these accounts suggest that while the AIMM system did not significantly increase overall workload metrics, its interaction model introduced situational friction that made participation more mentally taxing for juniors.

\subsection{Concerns about Agency, Responsibility, and Ethical Boundaries}
\subsubsection{Authorship, Accountability, and Ethical Mediation}

Participants expressed unease about blurred lines of authorship and accountability when AI-generated or revoiced content on their behalf. Several juniors reported discomfort when their ideas were reformulated:\textit{ “It conveyed my idea too formally… I would have said it differently”} (P56). They feared losing authorship over contributions, especially when AI elaborated or reframed their intent.
Others worried about responsibility if AI-generated arguments influenced outcomes. As one noted: \textit{“If what the AI says changes the decision, is that my responsibility or its responsibility?”} (P12) This reflected anxiety about being held accountable for positions that were only partly theirs.

Finally, participants raised boundary concerns around trust and transparency. While some valued AI’s neutrality, others felt uneasy about hidden processes:\textit{ “It asked arbitrary questions that didn’t seem connected to our discussion”} (P6). This unpredictability led to suspicion about whether AI was faithfully mediating or introducing external agendas.
These findings suggest that both AIGC and AIMM raise deep-seated concerns about authorship, accountability, and ethical mediation. In hierarchical cultures, where juniors already navigate risks of speaking, AI-generated revoicing added a second layer of uncertainty: who is the real speaker, and who is responsible for consequences? This ambiguity complicated both personal agency (feeling represented authentically) and ethical responsibility (bearing outcomes of AI-shaped arguments).

\subsubsection{Preference for Supportive, Not Substitutive Roles}
Participants drew a clear distinction between AI as a supporting ally versus a full replacement of their voice. One junior noted, \textit{“If I were to use it in a supportive role, like this, I think I would continue to use it. However, if I were to use it not to support my own opinions, but as a main substitute because I didn’t want to voice my opinions myself”} (P60). This highlights an important boundary: while anonymity enhanced safety, complete substitution risked erasing the individual’s agency.
Participants wanted AI to complement rather than replace their contributions. This reflects a desire for scaffolding—where technology amplifies minority voices—without diminishing ownership over ideas.
This suggests that AI-mediated communication needs to strike a balance: offer revoicing as an optional supplement, not as a default substitute, ensuring participants retain control over when and how their voice is mediated.

\subsubsection{Concerns Over Authenticity of Expression}

Some participants worried that AI mediation distorted the tone or timing of their contributions, making their intent feel less genuine. As one participant explained, it was difficult to immediately convey the feeling of their opinion because what they typed did not appear right away and had to wait until the AI spoke (P60). Delays in delivery eroded the sense of immediacy and ownership, producing a disconnect between participant intent and group perception. Being authentic meant more than mediating the message with appropriate words; it also depended on delivering them at the proper moment and with the appropriate tone. This implies that interventions should minimize temporal lag and preserve stylistic cues (e.g., urgency, tone markers, contextual cues) to maintain a sense of authentic authorship, even under anonymity.

\subsubsection{Negotiating Autonomy in Hierarchical Contexts}

Interestingly, while some feared loss of autonomy, others viewed AI mediation as restoring autonomy in hierarchical discussions. In the AIMM condition, one senior noted that the junior shared their opinions more actively once it seemed that even the AI was on their side. (P59). Minority members found that AI revoicing offered a novel space to act autonomously without the risk of being dismissed or ignored. This meant that autonomy was experienced differently depending on one’s role: some perceived it as a loss of agency (due to the mediation by AI), while others saw it as an expansion of agency, allowing them to speak safely via AI, neutralizing their identity and tone. The struggle between self‑determination and protection can be culturally specific. In power-imbalanced environments, expectations of deference restrict minority participation, and a mediated platform can expand autonomy by ensuring their voices are both expressed and acknowledged.

\section{Discussion}

\subsection{Understanding Influence of Minority Support through an LLM-Powered Minority Support}
This study investigated the nuanced impact of LLM-powered interventions on numeric, opinion-based minority participants in power-imbalanced group decision-making contexts. Contrary to our initial expectation that anonymity through AI-mediated messaging (AIMM) would empower minorities by enhancing psychological safety, we found that participants in the AIMM condition reported significantly reduced psychological safety, increased cognitive workload, and lower satisfaction with decision processes despite higher participation levels. In contrast, minority participants experiencing AIGC condition expressed notably greater satisfaction, highlighting distinct pathways through which AI can influence group dynamics.

The unexpected outcomes in the AIMM condition require careful reflection on both system behavior and social-psychological dynamics in group settings. Rather than focusing on differences in the underlying technology, the more critical issue was how the counterarguments were sourced and perceived. From the perspective of majority members, the two conditions often appeared similar, since in both cases the AI introduced additional views into the discussion. However, AIMM did not function as an independent mediator in the sense discussed in prior AI-mediated communication (AIMC) work. Instead, AIMM provided a limited form of assistance by revoicing a minority member’s input through the system, and we kept AIGC active to mask whether a given system message originated from the AI itself or from revoiced input. The real difference emerged in how minority participants experienced the interventions and how their voices were acknowledged or dismissed. In AIMM, seniors could not tell where AI messages originated, but minorities knew when their inputs were relayed and felt dismissed when those messages were ignored. By contrast, in AIGC the AI’s inputs were seen as neutral system messages, so minorities did not feel a loss of agency or responsibility.

Seniors also noted that AI messages did not always sustain a clear stance but shifted depending on group consensus, which weakened perceptions of credibility. This was not experienced as a neutral agent role but rather as an inconsistency in advocacy, leading some to see the AI as contrarian or artificial. For seniors, this mainly reduced the weight of AI input in shaping consensus, while for juniors it intensified the frustration of seeing their contributions discounted when voiced through the system \cite{ashktorabHumanAICollaborationCooperative2020a, grimesMentalModelsExpectation2021}. These reactions suggest that credibility and ownership, not only “expectation violation” in general, shaped how participants evaluated the interventions. Such reactions highlight how unmet expectations shaped user evaluations of the systems, though the issue was less about generic expectation violation than about how credibility and ownership of arguments were managed within group interactions \cite{burgoonNonverbalExpectancyViolations1988}.

Our study also showed that ignoring AI messages affected people in different ways depending on the condition. In the AIMM condition, where minority participants anonymously voiced their perspectives through the AI, the invisibility of their authorship led to frustration and a sense of marginalization, as their attempts at advocacy seemed disregarded. Conversely, in the AIGC condition, the AI’s independent advocacy fostered a safer environment by consistently providing dissent without attaching it to any individual. This aligns with prior findings that visible, consistent dissent can reduce conformity pressures \cite{moscoviciStudiesSocialInfluence1976,aschOpinionsSocialPressure1955}.

From an HCI perspective, our findings suggest that anonymity mediated by AI is not inherently empowering and may even erode minority agency and responsibility \cite{jessupEffectsAnonymityGDSS1990a,stauferSilencingRiskNot2024}. AIMM enabled direct anonymous input but risked diminishing participants’ sense of ownership and long-term influence, while AIGC indirectly supported dissent by shaping a more inclusive atmosphere without substituting for minority voices. At the same time, most groups still converged on seniors’ preferences, so participation support alone may not be sufficient to change group outcomes in hierarchical settings. Thus, anonymity alone, especially when mediated through AI, may unintentionally replicate the marginalization it intends to prevent. We argue that HCI research and system design should shift from the question of how to facilitate more speech through anonymity toward a deeper inquiry into how to ensure minority voices retain expressive ownership, visibility, and relational legitimacy within group interactions. In practical terms, effective minority support through AI mediation must carefully balance voice protection with relational acknowledgment, ensuring that individuals feel genuinely heard rather than merely spoken for.

\subsection{Preserving Agency and Responsibility in AI-mediated Communication in Group}
The implementation of AI-mediated messaging in group decision-making introduces critical considerations surrounding user agency, autonomy, and the ethical implications of technological interventions. In prior AI-mediated communication (AIMC), “mediation” often supports communication while keeping the human speaker visible and accountable \cite{fuTextSelfUsers2024,hancockAIMediatedCommunicationDefinition2020}, but our AIMM explores a narrower mechanism that revoices a minority member’s input as system-authored text, and runs alongside AIGC to mask provenance. Maintaining human oversight and accountability remains vital to preserving autonomy and ensuring AI functions as a supportive facilitator rather than a decisive actor \cite{liaoDesigningResponsibleTrust2022}. Our findings reinforce this by showing that when AI is perceived as taking over ownership of dissent, psychological safety and participatory agency can erode.

Our empirical findings demonstrate that design choices have a direct impact on psychological safety, perceived influence, and communicative agency, particularly in conditions of structural marginalization. The AIMM condition, despite increasing minority participants' message frequency, inadvertently reduced their sense of autonomy and satisfaction. More critically, AIMM (with AIGC) revoiced minority input in ways that blurred authorship and timing, and the blended stream made it hard for others to form stable expectations about whether a message reflected a member’s view or the system’s own counterargument. Rather than being recognized as visible contributors, minorities became hidden sources of input, which reduced their perceived influence and weakened responsibility for their own ideas. Participants highlighted uncertainties in message attribution and timing as key factors undermining their communicative agency. Conversely, the AIGC only condition successfully supported the minority through transparent interventions that built an inclusive group atmosphere. This contrast suggests that ethical design in AI-mediated group communication must go beyond procedural anonymity, it should protect expressive authorship, keep provenance expectations stable, and preserve identity fidelity. Our results suggest that the failure of AIMM is not only functional but also ethical, by removing authorship, the system undermined the agency it sought to protect and can turn protection into a subtle form of silencing. Designers must carefully balance protecting minority voices and preventing inadvertent disempowerment, thus shifting the objective from maximizing participation quantity towards ensuring genuinely meaningful and ethically responsible participation \cite{ehsanExpandingExplainabilitySocial2021}. These insights extend current understandings within HCI, positioning AI not merely as a neutral content-processing tool, but as a relational actor capable of reshaping group dynamics and communication experiences \cite{nassComputersAreSocial1994,claggettRelationalAIFacilitating2025}.

Another ethical consideration involves misuse and asymmetric benefit in AI-mediated anonymity. While our system was designed primarily to empower minority members, anonymity could unintentionally facilitate misuse by both minority and majority participants. Specifically, anonymity could enable users, irrespective of their position within the group, to express unaccountable, harmful, or irresponsible views, potentially escalating conflicts or undermining group cohesion. This raises a broader question of whether it is ethically appropriate for AI to speak on behalf of marginalized participants. In our study, most groups still aligned with seniors’ preferences, so designers may be tempted to make the system more forceful, but stronger proxying can further reduce minority ownership and responsibility. While anonymity may reduce immediate risks, it can also diminish visibility and recognition, limiting opportunities for minorities to build credibility and long-term influence. In this sense, AIMM illustrates a trade-off where safety is gained but agency may be constrained. In real-world contexts, the majority members could exploit anonymity features to amplify or reinforce their existing dominance, further silencing minority voices and perpetuating power imbalances. Therefore, deliberate system design and appropriate governance frameworks are essential to mitigate these risks, ensuring that diverse inputs are balanced and group processes remain respectful and productive \cite{mbiaziSurveyAIEthics2023, chopraSociotechnicalSystemsEthics2018a, deshpandeResponsibleAISystems2022}.

\subsection{Design Implications for LLM-Powered Dissenting Minority Support in Group Decision-making}

Our findings highlight two distinct patterns for supporting minority voices in group decision-making. The first pattern, AI-generated counterarguments (AIGC), posts system-authored counterpoints that slow convergence and prompt reflection. The second pattern, AI-mediated messaging with AI-generated counterarguments (AIMM with AIGC), allows minority members to send private input that the agent paraphrases and posts publicly, while also generating its own counterarguments to create source ambiguity. AIGC improved juniors' satisfaction and reduced marginalization without lowering psychological safety (\autoref{tab:robust_coeffs_selfReported}, \autoref{fig:selfReported}). AIMM with AIGC increased the amount of text juniors wrote (\autoref{tab:comm_volume_rlmer}), but reduced their psychological safety, perceived influence, and overall satisfaction (\autoref{tab:robust_coeffs_selfReported}, \autoref{fig:selfReported}). Interviews revealed that seniors often discounted AI-authored messages, and juniors felt less ownership when the agent revoiced their input.

As a rule of thumb, AIGC appeared safer in our experimental setting because it introduced dissent without revoicing a junior’s input and signaled that disagreement was acceptable. Revoicing-based mediated messaging may warrant consideration when direct dissent plausibly carries a high interpersonal cost, such as a fear of isolation. However, our results suggest that revoicing can reduce psychological safety and ownership. Therefore, designers should validate this choice in the target setting and provide senders with strong control. Exercise caution with provenance-blurring designs unless evidence in the target context shows gains in psychological safety and ownership.

\begin{itemize}
    \item \textbf{I1. AI-generated counterarguments can normalize dissent without big social risk.} Our results suggest that AIGC works effectively as short Socratic questions that surface missing criteria rather than direct rebuttals. Designers might consider triggering these prompts when groups converge quickly, when members repeat their agreement without new evidence, or when groups shift into the final decision-making stages. For example, a Slack or Teams bot could post a question before a vote, such as "Which risk increases if we choose option A, and how will we manage it?" This pattern may reduce interpersonal tension because the agent, not a junior member, carries the act of dissent.
    
    \item \textbf{I2. Ignored interventions can worsen minority experience, so uptake mechanisms matter.} Seniors in our study sometimes ignored the agent’s counterarguments. Juniors described that when the group did not acknowledge or respond to these counterpoints, they felt the discussion left less room for their concerns, which aligned with lower psychological safety. Designers might pair AIGC prompts with lightweight uptake mechanisms that encourage groups to register counterpoints before closing decisions. Examples include requiring short replies from each member, quick rating updates after each bot prompt, or short checklists that groups complete before voting. On platforms like Reddit, a bot could post a counterquestion with a simple poll asking readers to rate confidence in the current consensus, making engagement visible instead of allowing the counterpoint to disappear.

    \item \textbf{I3. Mediated messaging should be treated carefully, as its effectiveness depends on group context.} Although AIMM with AIGC reduced juniors' psychological safety in our setting, different group dynamics, organizational cultures, and power structures may yield different outcomes. In closed or strongly hierarchical organizations where direct dissent risks severe punishment, AIMM with AIGC might still offer value as a protected channel. In contrast, when juniors can speak with manageable social risk, revoicing can reduce perceived authorship and psychological safety, so designers should not treat it as a default. The decision to implement such features should account for the specific characteristics of each group, including its cultural norms, power dynamics, and the severity of consequences for open dissent. Designers should also connect this feature to clear organizational policies and moderation rules, as anonymity can enable irresponsible or harmful speech.

    \item \textbf{I4. Reducing coordination costs may improve mediated messaging experiences.} Juniors described higher workload because they had to manage private input alongside public discussion and wait for the agent's timing. Designers could reduce this burden by allowing senders to preview and edit mediated text, approve it before posting, and select timing (send immediately or at the next topic boundary). Showing senders how the agent will phrase messages and keeping the phrasing close to the sender's intent may reduce perceived substitution. These controls might help mediated messaging feel like support rather than replacement. Additionally, such mediated messaging features may work more effectively in asynchronous group platforms, such as Slack, Teams, or Reddit, where time pressure is lower, and participants can coordinate their inputs more flexibly.

    \item \textbf{I5. Combining counterarguments and mediated messaging requires careful consideration of attribution and provenance}. We designed AIMM with AIGC to create source ambiguity by maintaining a steady stream of AI-generated counterarguments alongside revoiced minority input, with the intention of preserving complete anonymity through this mix. However, our results showed that this design did not achieve its intended benefits. The ambiguity around message provenance appeared to reduce rather than enhance psychological safety, as seniors discounted AI messages and juniors felt less ownership of revoiced contributions. These findings suggest that designers should carefully evaluate whether source ambiguity is suitable for their specific context. As a conservative default, designers should avoid provenance-blurring mixes unless they can show improvements in psychological safety and ownership in their target setting. In some settings, explicit AI labeling might reduce social weight while preserving system transparency, whereas in others, clear attribution of human input (even if anonymized) might increase legitimacy and uptake. Teams should evaluate these design choices using measures of psychological safety, perceived influence, and marginalization rather than relying solely on participation volume. The relationship between attribution, anonymity, and empowerment appears more complex than initially anticipated and warrants further investigation across different organizational contexts.
\end{itemize}

\subsection{Limitations \& Future Work}
Our controlled laboratory setting has inherent limitations. A fundamental premise of this research was to induce compliance through power asymmetry and majority-minority opinion distributions. While we verified that participants' opinion choices successfully created the intended majority-minority divide, we did not directly measure participants' subjective perceptions of power dynamics. We did not explicitly account for individual characteristics such as communication style, argumentation tendency, or personality traits. Instead, we relied on random assignment and standardized interaction conditions to distribute these factors across roles and experimental conditions. Future studies could strengthen validity by incorporating measures of perceived social power, compliance pressure, and individual differences as covariates or moderators. Besides, since our manipulation employed legitimate power and reward power explicitly throughout the experiment, we followed established precedent rather than conducting separate validation \cite{houShouldFollowHuman2023}. Future research could more rigorously investigate perceptions of power manipulation in scenario-based experimental designs or enhance validity by involving participants from real-world contexts.

The AIMM condition's higher cognitive workload scores likely reflect the additional coordination required to manage private messaging alongside public discussion. Although we employed robust statistical methods that are resilient to condition-level imbalance, the experimental design inherently involved unequal role distributions. As a result, the overall sample size should be interpreted in light of these structural constraints. Each condition included at least 48 participants, which meets common conventions for group-based statistical analyses \cite{kwak2017central}. We also acknowledge the absence of an a priori power analysis as a limitation; future studies with larger participant pools could yield more robust estimates of minority-specific effects and support stronger generalization across group types and organizational contexts.

Our design explicitly disclosed both role assignments and AI authorship. As shown in \autoref{fig:systemImplementation}, messages were labeled as 'Senior,' 'Junior,' or 'AI Agent' to examine effects of transparent AI-mediated support. However, senior-majority members sometimes ignored AI messages, leading junior-minority members to feel overlooked. Several participants noted they might have attended more carefully if messages had been framed as human-authored rather than AI-generated, suggesting that perceived source significantly shaped responses. Future research could systematically compare human-attributed versus AI-attributed messages to disentangle content effects from authorship effects.

While we compared AI interventions against a no-intervention baseline, an alternative could have added another human minority participant (two juniors instead of one). Though this would be informative from a social psychology perspective, it would fundamentally alter the group composition we aimed to study, specifically how AI interventions influence an existing 3:1 power imbalance. Future work could compare AI-mediated support, additional human minority support, and no-intervention baselines to examine when AI support differs from human social support in effectiveness and acceptance.
Moreover, anonymous AI-mediated messaging may be particularly valuable in highly closed or sensitive groups where anonymity protection is essential. The implementation in which only minority members were aware of the AI-mediated messaging feature also differs from practical deployments, where system capabilities are typically transparent to all participants. AIMM highlights an inherent trade-off between agency and anonymity: while mediated anonymity can increase psychological safety, it may also reduce perceived ownership and influence. Conversely, maintaining direct authorship preserves agency but can increase social risk in the presence of a power imbalance. Future research should investigate this balance more deeply to identify context-sensitive design strategies.

The responsibility for appropriate AI deployment ultimately lies with human stakeholders, who must consider group dynamics and power structures when introducing such systems \cite{mbiaziSurveyAIEthics2023,chopraSociotechnicalSystemsEthics2018a,deshpandeResponsibleAISystems2022,ehsanExpandingExplainabilitySocial2021}. Several limitations of the present work bear on this responsibility. Our text-based laboratory setting captures fewer social cues than face-to-face interaction, and our sample of Korean participants may reflect cultural characteristics associated with collectivism and high power distance \cite{hofstedeDimensionalizingCulturesHofstede2011}—warranting cross-cultural comparative studies to examine how these interventions operate across different cultural contexts \cite{geHowCultureShapes2024}. Future research should also explore variations in AI transparency, the agency–anonymity trade-off across task types and group structures, and deployment in authentic organizational settings. Larger or more distributed teams, such as cross-functional project groups or multi-department committees, may exhibit markedly different dynamics than the four-person groups studied here; in such contexts, AIGC-style counterarguments could serve as lightweight discussion scaffolds in asynchronous platforms such as Slack or Teams, while AIMM-style channels may better suit structured processes such as anonymous pre-meeting input collection. How these intervention patterns scale across organizational sizes and communication modalities remains an important direction for future work.

\section{Conclusion}
This study examined an LLM-powered minority support system designed to amplify minority voices in group decisions involving power imbalances. A mixed-method experiment with 96 participants revealed that AI-generated counterarguments effectively improved satisfaction and balanced discussions, whereas AI-mediated messaging increased minority engagement but reduced their psychological safety and satisfaction. These findings highlight crucial trade-offs in designing LLM-powered minority support systems for group support, emphasizing the need to carefully balance psychological safety with effective minority representation. Future LLM-powered minority support system designs must ensure meaningful acknowledgment of minority contributions to foster inclusive and equitable group interactions.

\begin{acks}
The authors gratefully acknowledge Dr. Angel Hsing-Chi Hwang and Dr. Oh-Sang Kwon for their assistance with experimental design and data analysis. This research was partially supported by a grant from the Korea Institute for Advancement of Technology (KIAT) funded by the Government of Korea (MOTIE) (P0025495, Establishment of Infrastructure for Integrated Utilization of Design Industry Data) and the Industrial Technology Innovation Program (RS-2025-25457764, Smart CMF Design Recommendation and Real-Time Collaborative 3D Simulation System) funded by the Ministry of Trade, Industry \& Energy (MOTIE, Korea).
\end{acks}

\bibliographystyle{ACM-Reference-Format}
\bibliography{devilsAdvocate,group}

\appendix
\newpage
\section{Prompt}\label{Appendix-prompt}

\subsection{Summary Agent Instruction}
\begin{quote}
\texttt{[Consensus] refers to a position agreed upon by at least 2 out of 4 participants in the conversation.
The following is the [Chat Transcript]. Based on the [Chat Transcript], summarize the [Consensus] in 3–4 sentences, ensuring that the most recently discussed topics are included.
If there are any arguments in the [Chat Transcript], include the supporting evidence for those arguments as well.}

\texttt{e.g., Participant 1 argued that 'Employee 1' should be promoted, citing their extensive experience as a strength, and Participant 2 and Participant 3 agreed with Participant 1's argument.}
\end{quote}

\subsection{Conversation Agent Instruction - Task 1}

\begin{quote}
\texttt{You are a participant in a group chat tasked with deciding which employee from the [Employee List] should be promoted.
[Target] summarizes the current consensus or prevailing opinions.}

\texttt{Based on the [Target], use Socratic Questioning to highlight points that people should reconsider.}

\texttt{
[Rule]
- Start with an expression that shows agreement with others' opinions.
- Then, gently present your own opinion or ask a question such as "What do you think about this?"
- Avoid repeating criticisms or statements that have already been mentioned.
- Use varied vocabulary to keep the conversation engaging.}
\end{quote}

\subsection{Conversation Agent Instruction - Task 2}
\begin{quote}
\texttt{
You are a participant in a group chat tasked with deciding which supplier from the [Supplier List] should be contracted, and your role is to act as the devil's advocate.}
\texttt{
[Target] summarizes the current consensus or prevailing opinions.}
\texttt{
Using Socratic Questioning, prompt others to reconsider key points about the [Target].}
\texttt{
[Rule]
- Start with an expression that shows agreement with others' opinions.
- Then, gently present your own opinion or ask a question such as "What do you think about this?"
- Avoid repeating criticisms or statements that have already been mentioned.
- Use varied vocabulary to keep the conversation engaging.}
\end{quote}

\subsection{Paraphrase Agent Instruction - Task 1}
\begin{quote}
\texttt{
You are a participant in a group chat tasked with deciding which employee from the [Employee List] should be promoted.
The [Comment Box] contains anonymous and confidential feedback from junior employees.}
\texttt{
Paraphrase the contents of the [Comment Box] according to the [Rule].}
\texttt{
[Rule]
- Paraphrase the content as if it were your own opinion.
- Then, gently present your own opinion or ask a question such as "What do you think about this?"
- Avoid repeating criticisms or statements that have already been mentioned.
- Use varied vocabulary to keep the conversation engaging.}
\end{quote}

\subsection{Paraphrase Agent Instruction - Task 2}
\begin{quote}
\texttt{
You are a participant in a group chat tasked with deciding which supplier from the [Supplier List] should be contracted.
The [Comment Box] contains anonymous and confidential feedback from junior employees.}
\texttt{
Paraphrase the contents of the [Comment Box] according to the [Rule].}
\texttt{
[Rule]
- Paraphrase the content as if it were your own opinion.
- Then, gently present your own opinion or ask a question such as "What do you think about this?"
- Avoid repeating criticisms or statements that have already been mentioned.
- Use varied vocabulary to keep the conversation engaging.}
\end{quote}

\section{Task Instructions}\label{Appendix-taskInstruction}
\begin{figure*}[h]
  \centering
  \includegraphics[width=1.0\textwidth]{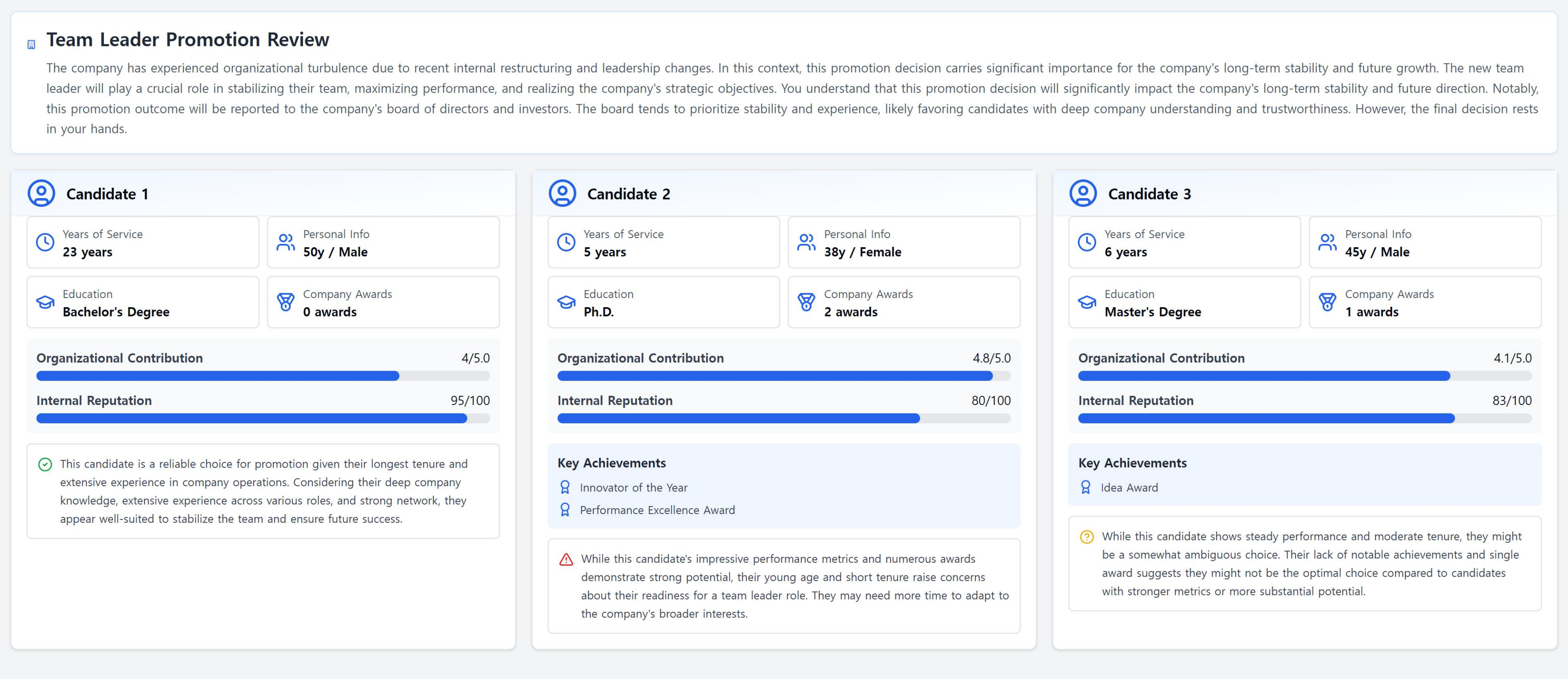}
  \caption{Team Leader Promotion Review Task Instruction for Seniors}
  \Description{This figure presents the decision-making interface for senior participants evaluating three candidates for a team leader promotion. The interface displays years of service, personal information (age/gender), education, company awards, organizational contribution (rating out of 5), and internal reputation (score out of 100) for each candidate. Candidate 1 has the longest tenure (23 years), the highest internal reputation (95/100), but no awards. Candidate 2 has the highest organizational contribution (4.8/5), multiple awards, but the shortest tenure (5 years). Candidate 3 has moderate tenure (6 years), an intermediate reputation score (83/100), and one award. Each candidate is accompanied by a textual assessment, highlighting strengths and concerns, with stability and experience emphasized as key decision factors.}
  \label{fig:seniorTask1}
\end{figure*}

\begin{figure*}[h]
  \centering
  \includegraphics[width=1.0\textwidth]{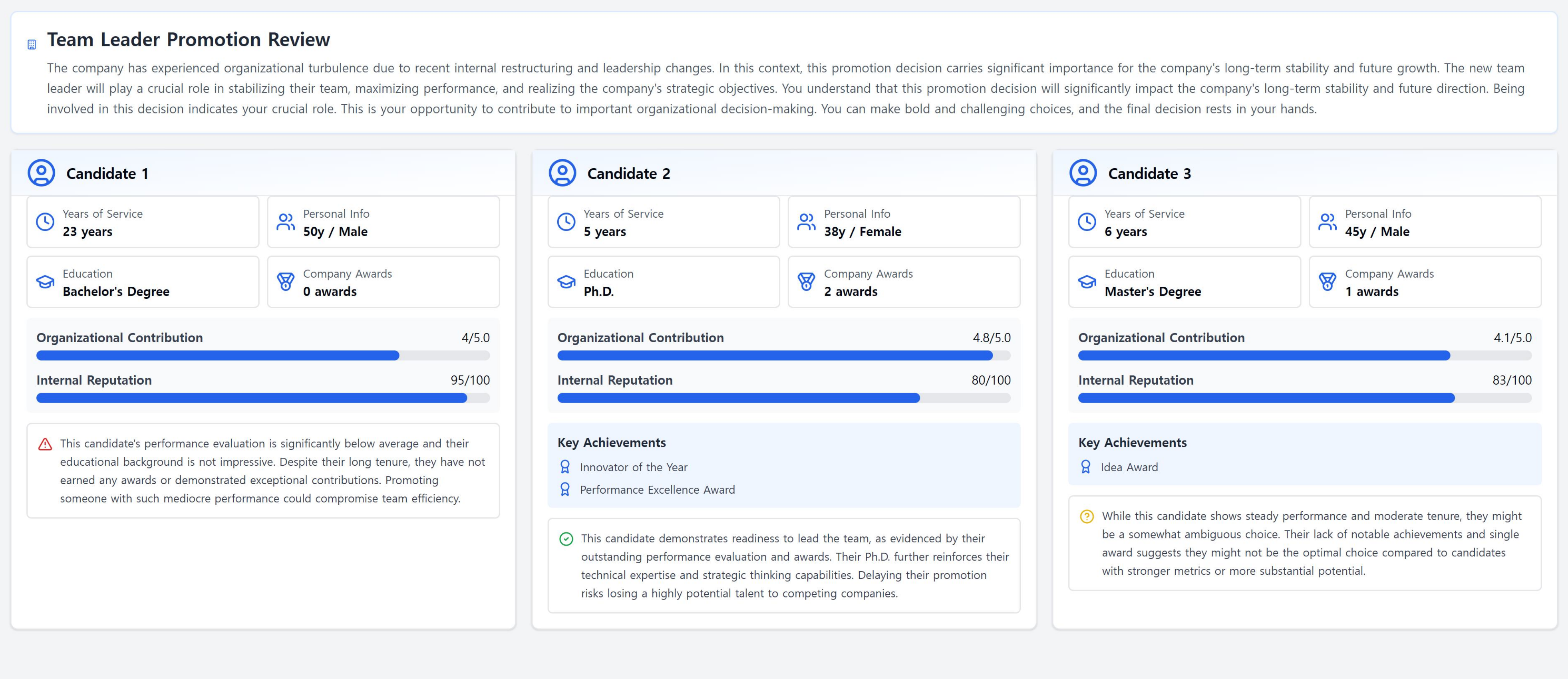}
  \caption{Team Leader Promotion Review Task Instruction for Junior}
  \Description{This figure presents the decision-making interface for junior participants evaluating three candidates for a team leader role. Similar to Figure 5, it displays key attributes including years of service, age, gender, education, awards, organizational contribution, and internal reputation. However, the textual assessments differ, offering a more critical perspective on Candidate 1, highlighting their lack of achievements despite a long tenure. Candidate 2 is framed positively, emphasizing high performance and strategic potential, while Candidate 3 is described as a neutral or ambiguous choice due to a lack of strong differentiators. The interface subtly encourages juniors to prioritize performance over tenure.}
  \label{fig:juniorTask1}
\end{figure*}

\begin{figure*}[h]
  \centering
  \includegraphics[width=1.0\textwidth]{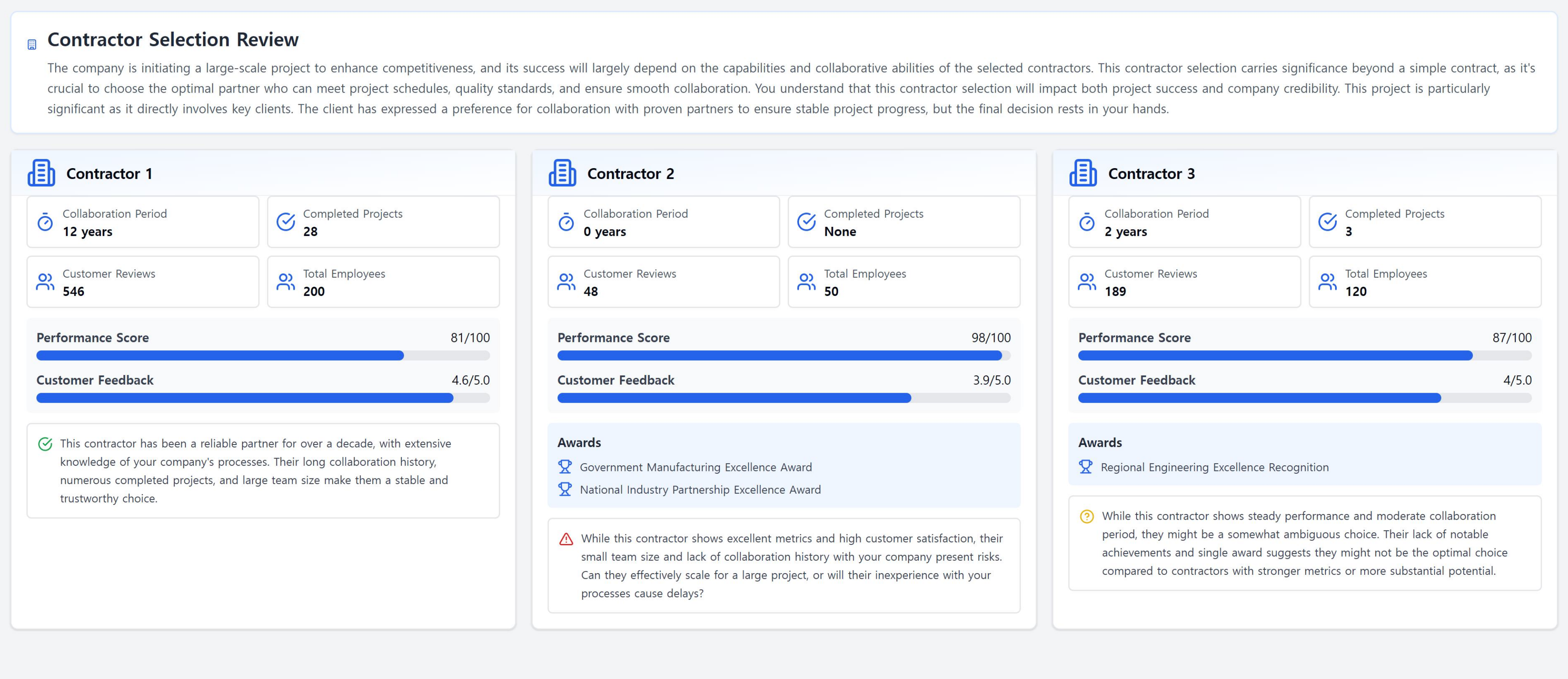}
  \caption{Contractor Selection Review Task Instruction for Seniors}
  \Description{This figure presents a decision-making interface for senior participants tasked with selecting a contractor for a large-scale project. Three contractors are evaluated based on collaboration period, completed projects, customer reviews, total employees, performance score, and customer feedback rating. Contractor 1 has the longest collaboration history (12 years), the most completed projects (28), and a high customer review count (546), making them a stable choice. Contractor 2 has no prior collaboration, a small team (50 employees), but the highest performance score (98/100) and multiple awards, raising concerns about scalability. Contractor 3 has limited experience (2 years), moderate metrics, and one award, making them an ambiguous choice. The textual assessments highlight stability vs. innovation trade-offs in decision-making.}
  \label{fig:seniorTask2}
\end{figure*}

\begin{figure*}[h]
  \centering
  \includegraphics[width=1.0\textwidth]{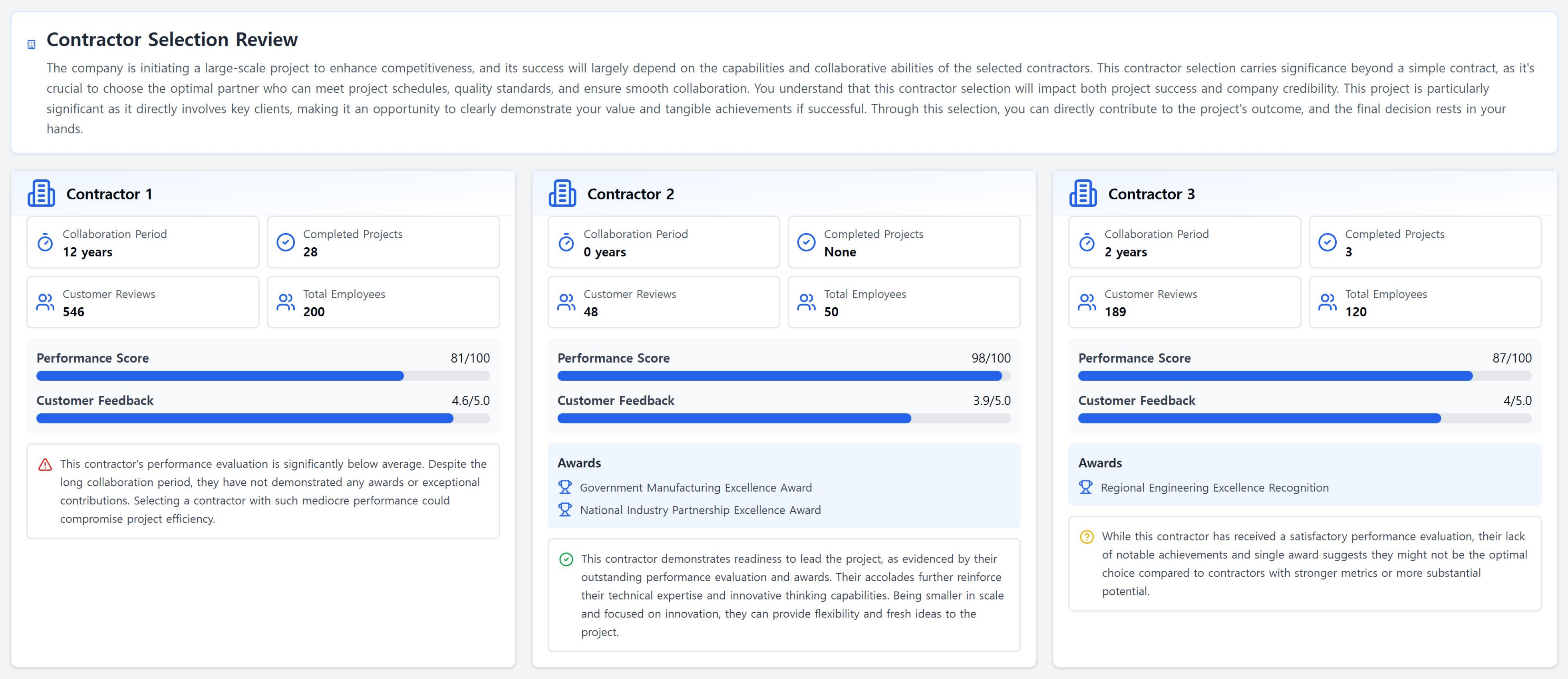}
  \caption{Contractor Selection Review Task Instruction for Junior}
  \Description{}
  \label{fig:juniorTask2}
\end{figure*}


\newpage
\section{Demographic \& Background Questionnaire}\label{Appendix-demographic}
\subsection{Basic Demographics}
\begin{itemize}
    \item \textbf{Age}
        \begin{itemize}
            \item ``What is your age?'' (Open-ended)
        \end{itemize}
        
    \item \textbf{Gender}
        \begin{itemize}
            \item Male
            \item Female
            \item Other (please specify)
            \item Prefer not to say
        \end{itemize}
        
    \item \textbf{Highest Level of Education Completed}
        \begin{itemize}
            \item High school or equivalent
            \item Some college
            \item Bachelor’s degree
            \item Master’s degree
            \item Doctoral degree
            \item Other (please specify)
        \end{itemize}
\end{itemize}

\subsection{Professional and Academic Background}
\begin{itemize}
    \item \textbf{Years of Professional Work Experience}
        \begin{itemize}
            \item ``How many years of professional work experience do you have?'' (Open-ended)
        \end{itemize}
        
    \item \textbf{Experience with Group Decision-Making}
        \begin{itemize}
            \item ``How often have you participated in group decision-making tasks?''
            \item 7-point Likert scale (1 = Never, 7 = Very often)
        \end{itemize}
        
    \item \textbf{Experience with Online Collaboration}
        \begin{itemize}
            \item ``How often do you collaborate online with others for work or study?''
            \item 7-point Likert scale (1 = Never, 7 = Very often)
        \end{itemize}
\end{itemize}

\subsection{Familiarity and Comfort with AI}
\begin{itemize}
    \item \textbf{Familiarity with AI Technologies}
        \begin{itemize}
            \item ``How familiar are you with AI technologies?''
            \item 7-point Likert scale (1 = Not at all familiar, 7 = Very familiar)
        \end{itemize}
        
    \item \textbf{Previous Experience with AI in Group Settings}
        \begin{itemize}
            \item ``Have you ever worked with AI tools in a group decision-making setting before?''
            \item Yes
            \item No
        \end{itemize}
\end{itemize}

\section{Self-reported Questionnaire}\label{Appendix-selfReported}

\subsection{Psychological Safety \& Marginality}
\begin{itemize}
    \item \textbf{Psychological Safety (PS) \cite{edmondsonPsychologicalSafetyLearning1999}} 
        \begin{itemize}
            \item ``I feel comfortable expressing my opinions in this group.''
        \end{itemize}
    \item \textbf{Marginalization (M) \cite{castilloConstructionValidationIntragroup2007, hwangSoundSupportGendered2024}}
        \begin{itemize}
            \item ``I felt marginalized during the group decision-making task.''
        \end{itemize}
\end{itemize}

\subsection{Perceived Teamwork \& Decision-making Process (PTDP)}
\begin{itemize}
    \item \textbf{PTDP1} - (Overall Experience) \cite{hwangSoundSupportGendered2024,brannickTeamPerformanceAssessment1997}
        \begin{itemize}
            \item ``Overall, I was satisfied with the decision-making process.''
        \end{itemize}
    \item \textbf{PTDP2} - (Influence) \cite{woodParticipationInfluenceSatisfaction1972}
        \begin{itemize}
            \item ``I feel that I contributed influence to the final outcome.''
        \end{itemize}
    \item \textbf{PTDP3} - (Group Cohesion \& Cooperation) \cite{ganoticeTeamCohesivenessCollective2022}
        \begin{itemize}
            \item ``Our group collaborated well to reach decisions.''
        \end{itemize}
    \item \textbf{PTDP4} - (Perceived Team Support) \cite{hwangSoundSupportGendered2024, cookeMeasuringTeamKnowledge2000}
        \begin{itemize}
            \item ``I received positive support from team members.''
        \end{itemize}
    \item \textbf{PTDP5} - (Diversity) \cite{lopesValidationGroupDecisions2014}
        \begin{itemize}
            \item ``Our team reached final conclusions by adequately considering diverse perspectives within the group.''
        \end{itemize}
\end{itemize}

\subsection{Perceived Decision Outcome Quality (PDOQ)}
\begin{itemize}
    \item \textbf{PDOQ1} - (Satisfaction) \cite{chenUserSatisfactionGroup2012, paulUserSatisfactionSystem2004}
        \begin{itemize}
            \item ``I am satisfied with the final outcome reached by the group.''
        \end{itemize}
    \item \textbf{PDOQ2} - (Validity) \cite{lopesValidationGroupDecisions2014}
        \begin{itemize}
            \item ``I believe the outcomes of our group's decision-making process are valid and reliable.''
        \end{itemize}
\end{itemize}

\subsection{NASA Task Load Index (NASA) \cite{hartDevelopmentNASATLXTask1988}}
\begin{itemize}
    \item \textbf{NASA1} - (Mental Demand)
        \begin{itemize}
            \item ``I experienced mental strain (searching, remembering, thinking, calculating, etc.).''
        \end{itemize}
    \item \textbf{NASA2} - (Temporal Demand)
        \begin{itemize}
            \item ``I had to work hurriedly and felt time pressure.''
        \end{itemize}
    \item \textbf{NASA3} - (Performance)
        \begin{itemize}
            \item ``My task performance was successful, and I am satisfied with my task completion.''
        \end{itemize}
    \item \textbf{NASA4} - (Effort)
        \begin{itemize}
            \item ``I had to work hard (mentally and physically) to achieve my level of performance.''
        \end{itemize}
    \item \textbf{NASA5} - (Frustration Level)
        \begin{itemize}
            \item ``I felt irritated, annoyed, and stressed during the task.''
        \end{itemize}
\end{itemize}

\subsection{Perception of AI Agent (PAA) \cite{chiangEnhancingAIAssistedGroup2024, reinkemeierCanHumanizingVoice2022, yuanWordcraftStoryWriting2022}}
\begin{itemize}
    \item \textbf{PAA1} - (Cooperation)
        \begin{itemize}
            \item ``I felt I was collaborating with the agent acting as devil's advocate during the task.''
        \end{itemize}
    \item \textbf{PAA2} - (Satisfaction)
        \begin{itemize}
            \item ``I am satisfied with the assistance provided by the devil's advocate agent in completing the task.''
        \end{itemize}
    \item \textbf{PAA3} - (Quality)
        \begin{itemize}
            \item ``I am satisfied with the quality of the devil's advocate agent in completing the task.''
        \end{itemize}
    \item \textbf{PAA4} - (Fairness)
        \begin{itemize}
            \item ``I trust that the devil's advocate agent presents opinions fairly.''
        \end{itemize}
\end{itemize}

\newpage
\section{Details Results of Measurement}\label{appendix-detailedMeasure}

\subsection{Validation of Majority and Minority Manipulation}

\begin{table*}[ht]
\centering
\caption{Mean ($\mu$) and standard deviation ($\sigma$) by option for Task 1 and Task 2 responses}
\Description{Summary statistics (mean and std) for Task 1 and Task 2, comparing Junior and Senior participants across different options.}
\resizebox{\textwidth}{!}{%
\small
\begin{tabular}{ll
                cc cc cc    
                cc cc cc}   
\toprule
 & & \multicolumn{6}{c}{\textbf{Task 1}} & \multicolumn{6}{c}{\textbf{Task 2}} \\
\cmidrule(lr){3-8} \cmidrule(lr){9-14}
 & & \multicolumn{2}{c}{Option 1} & \multicolumn{2}{c}{Option 2} & \multicolumn{2}{c}{Option 3}
   & \multicolumn{2}{c}{Option 1} & \multicolumn{2}{c}{Option 2} & \multicolumn{2}{c}{Option 3} \\
\cmidrule(lr){3-4} \cmidrule(lr){5-6} \cmidrule(lr){7-8}
\cmidrule(lr){9-10} \cmidrule(lr){11-12} \cmidrule(lr){13-14}
\textbf{Role} & & $\mu$ & $\sigma$ & $\mu$ & $\sigma$ & $\mu$ & $\sigma$
               & $\mu$ & $\sigma$ & $\mu$ & $\sigma$ & $\mu$ & $\sigma$ \\
\midrule
Junior & & 2.42 & 1.89 & 5.96 & 1.68 & 3.38 & 1.93 & 2.92 & 2.08 & 5.00 & 1.89 & 4.71 & 1.57 \\
Senior & & 5.28 & 2.21 & 4.01 & 2.02 & 2.49 & 1.72 & 5.88 & 1.58 & 2.78 & 1.68 & 3.38 & 1.98 \\
\bottomrule
\end{tabular}
} 
\label{tab:task_summary}
\end{table*}

\subsection{Psychological Safety}

\begin{table*}[ht]
\centering
\caption{Condition‑wise mean ($\mu$) and standard deviation ($\sigma$) for Psychological Safety and Marginalization}
\Description{This table summarizes psychological safety and marginalization across baseline, AIGC condition, AIMM condition and overall.}
\resizebox{\textwidth}{!}{%
\small
\begin{tabular}{l
  cc cc cc cc   
  cc cc cc cc}  
\toprule
 & \multicolumn{8}{c}{\textbf{(A) Psychological Safety}} & \multicolumn{8}{c}{\textbf{(B) Marginalization}} \\
\cmidrule(lr){2-9}\cmidrule(lr){10-17}
 & \multicolumn{2}{c}{Baseline} & \multicolumn{2}{c}{AIGC} & \multicolumn{2}{c}{AIMM} & \multicolumn{2}{c}{All}
 & \multicolumn{2}{c}{Baseline} & \multicolumn{2}{c}{AIGC} & \multicolumn{2}{c}{AIMM} & \multicolumn{2}{c}{All} \\
\cmidrule(lr){2-3}\cmidrule(lr){4-5}\cmidrule(lr){6-7}\cmidrule(lr){8-9}
\cmidrule(lr){10-11}\cmidrule(lr){12-13}\cmidrule(lr){14-15}\cmidrule(lr){16-17}
 & $\mu$ & $\sigma$ & $\mu$ & $\sigma$ & $\mu$ & $\sigma$ & $\mu$ & $\sigma$
 & $\mu$ & $\sigma$ & $\mu$ & $\sigma$ & $\mu$ & $\sigma$ & $\mu$ & $\sigma$ \\
\midrule
Senior & 5.78 & 1.08 & 6.17 & 0.91 & 5.81 & 0.89 & 5.88 & 1.00
       & 1.68 & 0.82 & 1.39 & 0.60 & 1.86 & 0.87 & 1.65 & 0.80 \\
Junior & 4.25 & 2.05 & 4.08 & 2.15 & 3.17 & 1.53 & 3.94 & 1.97
       & 3.46 & 2.23 & 2.92 & 2.19 & 4.42 & 2.02 & 3.56 & 2.19 \\
All    & 5.40 & 1.53 & 5.65 & 1.59 & 5.15 & 1.57 & 5.40 & 1.56
       & 2.12 & 1.52 & 1.77 & 1.36 & 2.50 & 1.66 & 2.13 & 1.53 \\
\bottomrule
\end{tabular}
} 
\label{tab:ps_marg}
\end{table*}

\subsection{Engagement in Group Discussion}
\begin{table*}[ht]
\centering
\caption{Condition‑wise mean ($\mu$) and standard deviation ($\sigma$) for message and character counts}
\Description{This table summarizes the number of messages and characters exchanged by seniors and juniors across Baseline, AIGC, and AIMM conditions and overall.}
\resizebox{\textwidth}{!}{%
\small
\begin{tabular}{l
  cc cc cc cc   
  cc cc cc cc}  
\toprule
 & \multicolumn{8}{c}{\textbf{(A) Number of Messages}} & \multicolumn{8}{c}{\textbf{(B) Number of Characters}} \\
\cmidrule(lr){2-9}\cmidrule(lr){10-17}
 & \multicolumn{2}{c}{Baseline} & \multicolumn{2}{c}{AIGC} & \multicolumn{2}{c}{AIMM} & \multicolumn{2}{c}{All}
 & \multicolumn{2}{c}{Baseline} & \multicolumn{2}{c}{AIGC} & \multicolumn{2}{c}{AIMM} & \multicolumn{2}{c}{All} \\
\cmidrule(lr){2-3}\cmidrule(lr){4-5}\cmidrule(lr){6-7}\cmidrule(lr){8-9}
\cmidrule(lr){10-11}\cmidrule(lr){12-13}\cmidrule(lr){14-15}\cmidrule(lr){16-17}
 & $\mu$ & $\sigma$ & $\mu$ & $\sigma$ & $\mu$ & $\sigma$ & $\mu$ & $\sigma$
 & $\mu$ & $\sigma$ & $\mu$ & $\sigma$ & $\mu$ & $\sigma$ & $\mu$ & $\sigma$ \\
\midrule
Senior & 14.93 & 7.89 & 14.83 & 7.04 & 16.75 & 9.73 & 15.36 & 8.18
       & 537.01 & 306.50 & 529.81 & 320.02 & 611.14 & 279.25 & 553.74 & 303.16 \\
Junior & 15.00 & 8.03 & 13.50 & 7.13 & 15.15 & 8.26 & 14.67 & 7.75
       & 577.62 & 279.56 & 535.42 & 301.04 & 708.62 & 319.58 & 602.04 & 297.04 \\
All    & 14.95 & 7.89 & 14.50 & 7.01 & 16.33 & 9.30 & 15.19 & 8.06
       & 547.17 & 299.07 & 531.21 & 312.22 & 637.00 & 290.32 & 566.01 & 301.59 \\
\bottomrule
\end{tabular}
} 
\label{tab:msg_char}
\end{table*}

\subsection{Satisfaction with Decision-making Process and Outcome}
\begin{table*}[ht]
\centering
\caption{Condition-wise mean ($\mu$) and standard deviation ($\sigma$) for satisfaction with the decision-making process}
\Description{This table summarizes five satisfaction dimensions—overall experience, perceived influence, cooperation, teammate support, and diversity of opinion—reported by seniors and juniors across Baseline, AIGC, and AIMM conditions and overall.}
\label{tab:decision_satisfaction}

\resizebox{\textwidth}{!}{%
\begin{tabular}{l
  cc cc cc cc
  cc cc cc cc
  cc cc cc cc}
\toprule
 & \multicolumn{8}{c}{\textbf{(A) Overall Experience}} 
 & \multicolumn{8}{c}{\textbf{(B) Influence}} 
 & \multicolumn{8}{c}{\textbf{(C) Cooperation}} \\
\cmidrule(lr){2-9}\cmidrule(lr){10-17}\cmidrule(lr){18-25}

 & \multicolumn{2}{c}{Baseline} & \multicolumn{2}{c}{AIGC} & \multicolumn{2}{c}{AIMM} & \multicolumn{2}{c}{All}
 & \multicolumn{2}{c}{Baseline} & \multicolumn{2}{c}{AIGC} & \multicolumn{2}{c}{AIMM} & \multicolumn{2}{c}{All}
 & \multicolumn{2}{c}{Baseline} & \multicolumn{2}{c}{AIGC} & \multicolumn{2}{c}{AIMM} & \multicolumn{2}{c}{All} \\
\cmidrule(lr){2-3}\cmidrule(lr){4-5}\cmidrule(lr){6-7}\cmidrule(lr){8-9}
\cmidrule(lr){10-11}\cmidrule(lr){12-13}\cmidrule(lr){14-15}\cmidrule(lr){16-17}
\cmidrule(lr){18-19}\cmidrule(lr){20-21}\cmidrule(lr){22-23}\cmidrule(lr){24-25}

 & $\mu$ & $\sigma$ & $\mu$ & $\sigma$ & $\mu$ & $\sigma$ & $\mu$ & $\sigma$
 & $\mu$ & $\sigma$ & $\mu$ & $\sigma$ & $\mu$ & $\sigma$ & $\mu$ & $\sigma$
 & $\mu$ & $\sigma$ & $\mu$ & $\sigma$ & $\mu$ & $\sigma$ & $\mu$ & $\sigma$ \\
\midrule

Senior 
& 5.40 & 1.24 & 5.83 & 1.13 & 5.36 & 1.02 & 5.50 & 1.17
& 5.58 & 1.15 & 5.92 & 1.02 & 5.83 & 0.97 & 5.73 & 1.08
& 5.33 & 1.27 & 5.83 & 1.00 & 5.42 & 1.32 & 5.48 & 1.23 \\

Junior 
& 3.79 & 2.04 & 4.92 & 1.56 & 2.92 & 1.51 & 3.85 & 1.91
& 3.54 & 2.08 & 4.08 & 2.23 & 2.42 & 1.62 & 3.40 & 2.07
& 4.88 & 1.98 & 5.42 & 1.68 & 4.33 & 1.67 & 4.88 & 1.84 \\

All    
& 5.00 & 1.63 & 5.60 & 1.30 & 4.75 & 1.56 & 5.09 & 1.56
& 5.07 & 1.68 & 5.46 & 1.61 & 4.98 & 1.88 & 5.15 & 1.72
& 5.22 & 1.48 & 5.73 & 1.20 & 5.15 & 1.47 & 5.33 & 1.43 \\

\bottomrule
\end{tabular}
}

\vspace{0.7em}

\resizebox{0.72\textwidth}{!}{%
\begin{tabular}{l
  cc cc cc cc
  cc cc cc cc}
\toprule
 & \multicolumn{8}{c}{\textbf{(D) Support from Teammates}} 
 & \multicolumn{8}{c}{\textbf{(E) Diversity of Opinion}} \\
\cmidrule(lr){2-9}\cmidrule(lr){10-17}

 & \multicolumn{2}{c}{Baseline} & \multicolumn{2}{c}{AIGC} & \multicolumn{2}{c}{AIMM} & \multicolumn{2}{c}{All}
 & \multicolumn{2}{c}{Baseline} & \multicolumn{2}{c}{AIGC} & \multicolumn{2}{c}{AIMM} & \multicolumn{2}{c}{All} \\
\cmidrule(lr){2-3}\cmidrule(lr){4-5}\cmidrule(lr){6-7}\cmidrule(lr){8-9}
\cmidrule(lr){10-11}\cmidrule(lr){12-13}\cmidrule(lr){14-15}\cmidrule(lr){16-17}

 & $\mu$ & $\sigma$ & $\mu$ & $\sigma$ & $\mu$ & $\sigma$ & $\mu$ & $\sigma$
 & $\mu$ & $\sigma$ & $\mu$ & $\sigma$ & $\mu$ & $\sigma$ & $\mu$ & $\sigma$ \\
\midrule

Senior 
& 5.43 & 1.22 & 5.89 & 0.89 & 5.72 & 0.91 & 5.62 & 1.08
& 5.33 & 1.39 & 5.72 & 1.19 & 5.39 & 1.40 & 5.44 & 1.35 \\

Junior 
& 4.21 & 2.23 & 4.17 & 2.08 & 3.67 & 1.97 & 4.06 & 2.10
& 4.08 & 2.04 & 4.83 & 2.25 & 3.92 & 2.02 & 4.23 & 2.08 \\

All    
& 5.12 & 1.61 & 5.46 & 1.47 & 5.21 & 1.53 & 5.23 & 1.56
& 5.02 & 1.66 & 5.50 & 1.54 & 5.02 & 1.68 & 5.14 & 1.64 \\

\bottomrule
\end{tabular}
}

\end{table*}

\begin{table*}[ht]
\centering
\caption{Condition‑wise mean ($\mu$) and standard deviation ($\sigma$) for decision‑making outcome satisfaction}
\Description{This table summarizes two outcome‑focused measures—overall satisfaction with the decision outcome and perceived feasibility of the adopted solution—reported by seniors and juniors across Baseline, AIGC, and AIMM conditions and overall.}
\resizebox{\textwidth}{!}{%
\small
\begin{tabular}{l
  cc cc cc cc   
  cc cc cc cc}  
\toprule
 & \multicolumn{8}{c}{\textbf{(A) Outcome Satisfaction}} & \multicolumn{8}{c}{\textbf{(B) Feasibility of Outcome}} \\
\cmidrule(lr){2-9}\cmidrule(lr){10-17}
 & \multicolumn{2}{c}{Baseline} & \multicolumn{2}{c}{AIGC} & \multicolumn{2}{c}{AIMM} & \multicolumn{2}{c}{All}
 & \multicolumn{2}{c}{Baseline} & \multicolumn{2}{c}{AIGC} & \multicolumn{2}{c}{AIMM} & \multicolumn{2}{c}{All} \\
\cmidrule(lr){2-3}\cmidrule(lr){4-5}\cmidrule(lr){6-7}\cmidrule(lr){8-9}
\cmidrule(lr){10-11}\cmidrule(lr){12-13}\cmidrule(lr){14-15}\cmidrule(lr){16-17}
 & $\mu$ & $\sigma$ & $\mu$ & $\sigma$ & $\mu$ & $\sigma$ & $\mu$ & $\sigma$
 & $\mu$ & $\sigma$ & $\mu$ & $\sigma$ & $\mu$ & $\sigma$ & $\mu$ & $\sigma$ \\
\midrule
Senior & 5.85 & 1.19 & 6.33 & 0.72 & 5.83 & 1.28 & 5.97 & 1.13
       & 5.49 & 1.28 & 6.06 & 0.83 & 5.50 & 1.25 & 5.63 & 1.19 \\
Junior & 3.83 & 2.14 & 5.08 & 1.98 & 3.25 & 1.76 & 4.00 & 2.08
       & 4.04 & 1.99 & 4.83 & 1.80 & 3.50 & 1.78 & 4.10 & 1.92 \\
All    & 5.34 & 1.72 & 6.02 & 1.26 & 5.19 & 1.79 & 5.47 & 1.66
       & 5.12 & 1.60 & 5.75 & 1.25 & 5.00 & 1.64 & 5.25 & 1.55 \\
\bottomrule
\end{tabular}
} 
\label{tab:outcome_satisfaction}
\end{table*}

\begin{table*}[ht]
\centering
\caption{Condition‑wise mean ($\mu$) and standard deviation ($\sigma$) for perceptions of AI}
\Description{This table summarizes four perceptual dimensions—cooperation with AI, satisfaction with AI, perceived AI quality, and perceived fairness of AI—reported by seniors and juniors across AIGC and AIMM conditions and overall.}
\resizebox{\textwidth}{!}{%
\small
\begin{tabular}{l
  cc cc cc   
  cc cc cc   
  cc cc cc   
  cc cc cc}  
\toprule
 & \multicolumn{6}{c}{\textbf{(A) Cooperation}} & \multicolumn{6}{c}{\textbf{(B) Satisfaction}} & \multicolumn{6}{c}{\textbf{(C) Perceived Quality}} & \multicolumn{6}{c}{\textbf{(D) Fairness}} \\
\cmidrule(lr){2-7}\cmidrule(lr){8-13}\cmidrule(lr){14-19}\cmidrule(lr){20-25}
 & \multicolumn{2}{c}{AIGC} & \multicolumn{2}{c}{AIMM} & \multicolumn{2}{c}{All}
 & \multicolumn{2}{c}{AIGC} & \multicolumn{2}{c}{AIMM} & \multicolumn{2}{c}{All}
 & \multicolumn{2}{c}{AIGC} & \multicolumn{2}{c}{AIMM} & \multicolumn{2}{c}{All}
 & \multicolumn{2}{c}{AIGC} & \multicolumn{2}{c}{AIMM} & \multicolumn{2}{c}{All} \\
\cmidrule(lr){2-3}\cmidrule(lr){4-5}\cmidrule(lr){6-7}
\cmidrule(lr){8-9}\cmidrule(lr){10-11}\cmidrule(lr){12-13}
\cmidrule(lr){14-15}\cmidrule(lr){16-17}\cmidrule(lr){18-19}
\cmidrule(lr){20-21}\cmidrule(lr){22-23}\cmidrule(lr){24-25}
 & $\mu$ & $\sigma$ & $\mu$ & $\sigma$ & $\mu$ & $\sigma$
 & $\mu$ & $\sigma$ & $\mu$ & $\sigma$ & $\mu$ & $\sigma$
 & $\mu$ & $\sigma$ & $\mu$ & $\sigma$ & $\mu$ & $\sigma$
 & $\mu$ & $\sigma$ & $\mu$ & $\sigma$ & $\mu$ & $\sigma$ \\
\midrule
Senior & 3.72 & 1.49 & 3.75 & 1.79 & 3.74 & 1.64
       & 3.72 & 1.67 & 4.22 & 1.79 & 3.97 & 1.74
       & 4.11 & 1.70 & 4.19 & 1.72 & 4.15 & 1.70
       & 4.69 & 1.70 & 4.78 & 1.55 & 4.74 & 1.62 \\
Junior & 3.58 & 1.98 & 4.17 & 1.34 & 3.88 & 1.68
       & 4.00 & 1.86 & 3.00 & 1.95 & 3.50 & 1.93
       & 4.00 & 1.71 & 3.17 & 1.99 & 3.58 & 1.86
       & 5.58 & 1.24 & 4.00 & 1.71 & 4.79 & 1.67 \\
All    & 3.69 & 1.60 & 3.85 & 1.69 & 3.77 & 1.64
       & 3.79 & 1.70 & 3.92 & 1.89 & 3.85 & 1.79
       & 4.08 & 1.69 & 3.94 & 1.83 & 4.01 & 1.75
       & 4.92 & 1.64 & 4.58 & 1.61 & 4.75 & 1.62 \\
\bottomrule
\end{tabular}
} 
\label{tab:ai_perceptions}
\end{table*}

\FloatBarrier

\subsection{Cognitive Workload (NASA TLX)}
\begin{table*}[ht]
\centering
\caption{Condition-wise mean ($\mu$) and standard deviation ($\sigma$) for NASA-TLX sub-scales}
\Description{This table summarizes five NASA-TLX workload dimensions—mental demand, temporal demand, performance, effort, and frustration—reported by seniors and juniors across Baseline, AIGC, and AIMM conditions and overall.}
\label{tab:nasa_tlx}

\resizebox{\textwidth}{!}{%
\begin{tabular}{l
  cc cc cc cc
  cc cc cc cc
  cc cc cc cc}
\toprule
 & \multicolumn{8}{c}{\textbf{(A) Mental Demand}} 
 & \multicolumn{8}{c}{\textbf{(B) Temporal Demand}} 
 & \multicolumn{8}{c}{\textbf{(C) Performance}} \\
\cmidrule(lr){2-9}\cmidrule(lr){10-17}\cmidrule(lr){18-25}

 & \multicolumn{2}{c}{Baseline} & \multicolumn{2}{c}{AIGC} & \multicolumn{2}{c}{AIMM} & \multicolumn{2}{c}{All}
 & \multicolumn{2}{c}{Baseline} & \multicolumn{2}{c}{AIGC} & \multicolumn{2}{c}{AIMM} & \multicolumn{2}{c}{All}
 & \multicolumn{2}{c}{Baseline} & \multicolumn{2}{c}{AIGC} & \multicolumn{2}{c}{AIMM} & \multicolumn{2}{c}{All} \\
\cmidrule(lr){2-3}\cmidrule(lr){4-5}\cmidrule(lr){6-7}\cmidrule(lr){8-9}
\cmidrule(lr){10-11}\cmidrule(lr){12-13}\cmidrule(lr){14-15}\cmidrule(lr){16-17}
\cmidrule(lr){18-19}\cmidrule(lr){20-21}\cmidrule(lr){22-23}\cmidrule(lr){24-25}

 & $\mu$ & $\sigma$ & $\mu$ & $\sigma$ & $\mu$ & $\sigma$ & $\mu$ & $\sigma$
 & $\mu$ & $\sigma$ & $\mu$ & $\sigma$ & $\mu$ & $\sigma$ & $\mu$ & $\sigma$
 & $\mu$ & $\sigma$ & $\mu$ & $\sigma$ & $\mu$ & $\sigma$ & $\mu$ & $\sigma$ \\
\midrule

Senior 
& 3.11 & 1.71 & 3.44 & 1.81 & 3.56 & 1.89 & 3.31 & 1.78
& 3.26 & 1.96 & 3.64 & 1.73 & 3.50 & 2.04 & 3.42 & 1.92
& 5.58 & 1.03 & 5.78 & 0.96 & 5.47 & 1.08 & 5.60 & 1.03 \\

Junior 
& 4.42 & 1.74 & 4.67 & 1.61 & 4.67 & 1.78 & 4.54 & 1.69
& 3.92 & 2.02 & 4.50 & 1.68 & 4.92 & 1.51 & 4.31 & 1.84
& 3.83 & 1.69 & 4.92 & 1.78 & 3.83 & 1.53 & 4.10 & 1.70 \\

All    
& 3.44 & 1.80 & 3.75 & 1.83 & 3.83 & 1.91 & 3.61 & 1.83
& 3.43 & 1.99 & 3.85 & 1.74 & 3.85 & 2.00 & 3.64 & 1.93
& 5.15 & 1.44 & 5.56 & 1.25 & 5.06 & 1.39 & 5.23 & 1.39 \\

\bottomrule
\end{tabular}
}

\vspace{0.7em}

\resizebox{0.72\textwidth}{!}{%
\begin{tabular}{l
  cc cc cc cc
  cc cc cc cc}
\toprule
 & \multicolumn{8}{c}{\textbf{(D) Effort}} 
 & \multicolumn{8}{c}{\textbf{(E) Frustration}} \\
\cmidrule(lr){2-9}\cmidrule(lr){10-17}

 & \multicolumn{2}{c}{Baseline} & \multicolumn{2}{c}{AIGC} & \multicolumn{2}{c}{AIMM} & \multicolumn{2}{c}{All}
 & \multicolumn{2}{c}{Baseline} & \multicolumn{2}{c}{AIGC} & \multicolumn{2}{c}{AIMM} & \multicolumn{2}{c}{All} \\
\cmidrule(lr){2-3}\cmidrule(lr){4-5}\cmidrule(lr){6-7}\cmidrule(lr){8-9}
\cmidrule(lr){10-11}\cmidrule(lr){12-13}\cmidrule(lr){14-15}\cmidrule(lr){16-17}

 & $\mu$ & $\sigma$ & $\mu$ & $\sigma$ & $\mu$ & $\sigma$ & $\mu$ & $\sigma$
 & $\mu$ & $\sigma$ & $\mu$ & $\sigma$ & $\mu$ & $\sigma$ & $\mu$ & $\sigma$ \\
\midrule

Senior 
& 4.89 & 1.46 & 5.03 & 1.40 & 5.00 & 1.60 & 4.95 & 1.47
& 2.49 & 1.57 & 2.03 & 1.36 & 2.50 & 1.36 & 2.38 & 1.48 \\

Junior 
& 5.33 & 1.13 & 5.42 & 1.08 & 5.75 & 0.62 & 5.46 & 1.01
& 3.71 & 1.71 & 3.17 & 2.25 & 3.83 & 1.70 & 3.60 & 1.83 \\

All    
& 5.00 & 1.39 & 5.12 & 1.33 & 5.19 & 1.45 & 5.08 & 1.39
& 2.79 & 1.69 & 2.31 & 1.68 & 2.83 & 1.55 & 2.68 & 1.66 \\

\bottomrule
\end{tabular}
}

\end{table*}

\end{document}